\documentclass[10pt,journal,compsoc]{IEEEtran}
\pdfoutput=1
\usepackage{fancyhdr}
\usepackage{graphicx}
\usepackage[normalem]{ulem}
\usepackage{hyperref}
\usepackage{amsmath}
\usepackage{booktabs} % For formal tables
\usepackage{upgreek}
\usepackage{graphicx}
\usepackage{verbatim}
\usepackage{multirow}
\usepackage{caption}
\usepackage{color}
\usepackage{array}
\usepackage{subfigure}
\usepackage{float}
\usepackage{CJK}
\usepackage{setspace}
\usepackage{caption}
\usepackage[]{footmisc}
\usepackage{tablefootnote}
\usepackage{indentfirst}
\usepackage{dblfloatfix} 
\usepackage{multirow}
\usepackage[utf8]{inputenc}
\usepackage[english]{babel}
\usepackage{amsthm}
\usepackage{setspace}
\usepackage{epstopdf}

\newtheorem{theorem}{Theorem}[section]
\newtheorem{corollary}{Corollary}[theorem]
\newtheorem{lemma}[theorem]{Lemma}

\graphicspath{{./figures/}}

\usepackage{colortbl}
\definecolor{mygray}{gray}{.9}
\definecolor{mypink}{rgb}{.99,.91,.95}
\definecolor{mycyan}{cmyk}{.3,0,0,0}

\usepackage{comment}
\usepackage{enumitem}
\usepackage[ruled,noline,noend]{algorithm2e}%vlined

\newcommand{\enquote}[1]{``#1''}

\newcommand{\CL} {\mathit{CL}}
\newcommand{\ML} {\mathit{ML}}
\newcommand{\GL} {\mathit{GL}}
\newcommand{\GW} {\mathit{GW}}
\newcommand{\GN} {\mathit{GN}}
\newcommand{\VGN} {\mathit{VGN}}

\newcommand{\CW} {\mathit{CW}}
\newcommand{\MW} {\mathit{MW}}
\newcommand{\CS} {\mathit{CS}}
\newcommand{\MS} {\mathit{MS}}
\newcommand{\GR} {\mathit{GR}}
\usepackage{mathabx}
\newcommand{\CRup} {\widehat{\mathit{CR}}}
\newcommand{\CLup} {\widehat{\mathit{CL}}}
\newcommand{\MRup} {\widehat{\mathit{MR}}}
\newcommand{\MLup} {\widehat{\mathit{ML}}}
\newcommand{\GRup} {\widehat{\mathit{GR}}}
\newcommand{\GLup} {\widehat{\mathit{GL}}}
\newcommand{\GWup} {\widehat{\mathit{GW}}}
\newcommand{\Sup} {\widehat{\mathit{S}}}
\newcommand{\Lup} {\widehat{\mathit{L}}}

\newcommand{\CLlo} {\widecheck{\mathit{CL}}}
\newcommand{\MRlo} {\widecheck{\mathit{MR}}}
\newcommand{\MLlo} {\widecheck{\mathit{ML}}}
\newcommand{\GRlo} {\widecheck{\mathit{GR}}}

\newcommand{\GWlo} {\widecheck{\mathit{GW}}}
\newcommand{\Slo} {\widecheck{\mathit{S}}}
\newcommand{\Rup} {\widehat{\mathit{R}}}
\newcommand{\RupA} {\widehat{\mathit{R1}}}
\newcommand{\RupB} {\widehat{\mathit{R2}}}

\ifCLASSOPTIONcompsoc
  \usepackage[nocompress]{cite}
\else
  \usepackage{cite}
\fi

\ifCLASSINFOpdf
\else
\fi

\begin{document}
\title{RTGPU: Real-Time GPU Scheduling of Hard Deadline Parallel Tasks with Fine-Grain Utilization}

\author{An~Zou,~\IEEEmembership{Member,~IEEE,}
        Jing~Li,~\IEEEmembership{Member,~IEEE,}\\
        Christopher~D.~Gill,~\IEEEmembership{Senior Member,~IEEE,}
        Xuan~Zhang,~\IEEEmembership{Member,~IEEE,}
        % <-this % stops a space
%\IEEEcompsocitemizethanks{\IEEEcompsocthanksitem An Zou is with Shanghai Jiao Tong University.\protect\\
%\IEEEcompsocthanksitem Jing Li is with New Jersey Institute of Technology.\protect\\
%\IEEEcompsocthanksitem Christopher D. Gill, and Xuan Zhang are with Washington University in St. Louis, St. Louis.
%MO, 63130.}% <-this % stops an unwanted space
\thanks{This work is supported by NSF CNS-1739643, NSF CNS-1948457, and NSFC 62202287.}\vspace{-2mm}}

\markboth{Published in IEEE Transactions on Parallel and Distributed Systems 2023. DOI: 10.1109/TPDS.2023.3235439}%
{Shell \MakeLowercase{\textit{et al.}}: Bare Demo of IEEEtran.cls for Computer Society Journals}

\IEEEtitleabstractindextext{%
\begin{abstract}
Many emerging cyber-physical systems, such as autonomous vehicles and robots, rely heavily on artificial intelligence and machine learning algorithms to perform important system operations. Since these highly parallel applications are computationally intensive, they need to be accelerated by graphics processing units (GPUs) to meet stringent timing constraints. However, despite the wide adoption of GPUs, efficiently scheduling multiple GPU applications while providing rigorous real-time guarantees remains a challenge. Each GPU application has multiple CPU execution and memory copy segments, with GPU kernels running on different hardware resources. Because of the complicated interactions between heterogeneous segments of parallel tasks, high schedulability is hard to achieve with conventional approaches. This paper proposes RTGPU, which combines fine-grain GPU partitioning on the system side with a novel scheduling algorithm on the theory side. Through system and theory co-design, RTGPU achieves superior system throughput and real-time schedulability. In this paper, we start by building a model for the CPU and memory copy segments. Leveraging persistent threads, we then implement fine-grained GPU partitioning with improved performance through interleaved execution. To reap the benefits of fine-grained GPU partitioning and schedule multiple parallel GPU applications, we propose a novel real-time scheduling algorithm based on federated scheduling and grid search with uniprocessor fixed-priority scheduling.  Our approach provides real-time guarantees to meet hard deadlines, and achieves over 11\% improvement in system throughput and up to 57\% schedulability improvement compared with previous work. We validate and evaluate RTGPU on NVIDIA GTX1080Ti GPU systems. Our system side techniques can be applied on mainstream NVIDIA GPUs, and the proposed scheduling theory can be used in general heterogeneous computing platforms which have a similar task execution pattern.
\end{abstract}

}

\maketitle

\IEEEdisplaynontitleabstractindextext

% \fi
\IEEEpeerreviewmaketitle

\IEEEraisesectionheading{\section{Introduction}\label{sec:intro}}
\IEEEPARstart{N}{owadays, }
artificial intelligence (AI) and machine learning (ML) applications accelerated by graphics processing units (GPUs) are widely adopted in emerging autonomous systems, such as self-driving vehicles and collaborative robotics~\cite{bojarski2016end,lin2018architectural}. For example, Volvo deployed NVIDIA DRIVE PX 2 technology for semi-autonomous driving in 100 XC90 luxury SUVs~\cite{NVIDIA_Volvo}. These autonomous systems need to  execute different AI/ML applications simultaneously in the GPU to perform tasks such as object detection, 3D annotation, movement prediction, and route planning~\cite{jafari2014real,yolov3}. Moreover, they often need to process images and signals from various sensors and decide the next action in real time.
It is thus essential to manage concurrent execution in the GPUs diligently with respect to various timing constraints, since they can have direct and critical impacts on the stability and safety of the whole system.

For general-purpose computing in a non-real-time setting, GPU scheduling aims to minimize the makespan of a single application or to maximize the total throughput of the system~\cite{rossbach2011ptask,kato2012gdev,kayiran2014managing,yang2011hybrid}. Many state-of-the-art learning frameworks that support GPU acceleration of AI and ML algorithms, such as Caffe~\cite{Caffe} and TensorFlow~\cite{abadi2016tensorflow}, also handle workloads in a sequential manner.
This type of sequential execution model is sufficient for large-scale resource-abundant systems (e.g., in data center applications) that aim to maximize the average throughput of a single task. The same model, however, does not translate well in parallel GPU applications with strict timing deadlines.
When computing resources are constrained, such as in on-board GPU systems, parallel tasks have to make good use of the limited resource to meet strict deadlines. However, even state-of-art GPU execution patterns pose unique challenges to the real-time scheduling of parallel tasks. First, inside each task, there are multiple serially dependent segments, namely, CPU execution and memory copy segments and GPU kernels. Inside each task, these segments need to access different hardware resources serially. Second, a large GPU kernel in one task may occupy the entire GPU, blocking the GPU kernels in other parallel tasks. This aggravated dependency inside and among tasks may reduce the system’s performance or cause extra scheduling pessimism under hard timing constraints.

This paper proposes RTGPU, a general real-time GPU scheduling framework. To overcome such aggravated dependency and pessimism in GPU real-time scheduling, RTGPU provides GPU partitioning and modeling as well as a scheduling algorithm and tight schedulability analysis. First, based on an in-depth understanding of GPU kernel execution and profiling of synthetic workloads, we leverage the persistent threads technique \cite{gupta2012study} to support Streaming Multiprocessor (SM)-granularity partitioning for concurrent GPU applications.
To fully utilize the GPU resources, we further propose interleaved execution which can achieve 10\% to 37\% improvement in system utilization compared with SM-granularity resource partitioning without interleaved execution \cite{liang2014efficient}.
We then develop a measurement-based task model that introduces the concept of \emph{virtual streaming multiprocessors} (virtual SMs), which allows finer-grained (SM-level) GPU scheduling without any low-level modifications to GPU systems. 

Following the flexible task execution pattern, we propose a novel real-time scheduling algorithm leveraging federated and fixed-priority scheduling. The key idea behind federated scheduling is to calculate and statically assign the specific computing resources that each parallel real-time task needs to meet its deadline. Note that preemption between tasks is not required if the correct number of fixed-granularity computing resources can be accurately derived in analysis and enforced during runtime. For the CPU segments and memory copies between CPU and GPU scheduled by the uniprocessor fixed-priority scheduling algorithm, a novel analysis is proposed to calculate the response time upper bounds and lower bounds of the two types of segments alternately. Leveraging the flexibility from GPU partitioning and the scheduling algorithm with tight response time analysis,
the proposed RTGPU achieves up to 57\% improvement in system schedulability. More generally, our proposed scheduling algorithm and analysis can be applied to other heterogeneous computing systems that have a similar application execution pattern (each task has CPU, memory copy, and heterogeneous core segments), such as AMD GPUs and Google TPUs.
\vspace{-4mm}
\section{Background}
\label{background}

%In this section we present background information on which our proposed approach builds.
%Section~\ref{sec:GPUbackground} gives an overview of how programs execute on GPUs.  
%Section~\ref{sec:persistentthreads} then introduces the persistent threads approach which we leverage in our work.
%Finally, Section~\ref{sec:suspension} describes the multi-segment self-suspension model on which our new scheduling approach depends.
%\vspace{-2mm}

\subsection{Background on GPU Systems}
\label{sec:GPUbackground}
\begin{figure}
\centering
\includegraphics[width=0.28\textwidth]{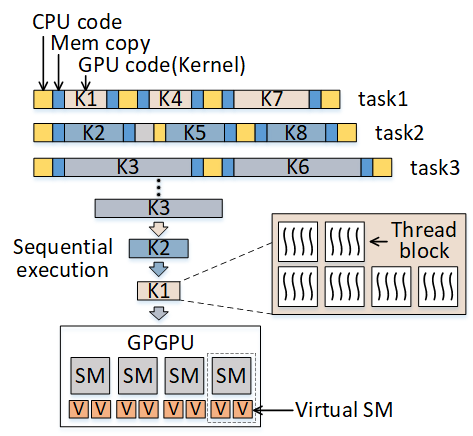}
\caption{Typical GPU task execution pattern.}
\label{fig:diagram0.jpg}
\vspace{-6mm}
\end{figure}

GPUs are designed to accelerate compute-intensive workloads with high levels of data parallelism. As shown in Fig.~\ref{fig:diagram0.jpg}., a typical GPU program contains three parts — a code segment that runs on the host CPU (the \emph{CPU segment}), the host/device \emph{memory copy segment}, and the device code segment which is also known as the \emph{GPU kernel}. GPU kernels are single instruction multiple threads (SIMT) programs. The programmer writes code for one thread, many threads are grouped into one thread block, and many thread blocks form a GPU kernel. The threads in one block execute the same instruction on different data simultaneously. A GPU consists of multiple streaming multiprocessors (SMs). The SM is the main computing unit, and each thread block is assigned to an SM to execute. Inside each SM are many smaller execution units that handle the physical execution of the threads in a thread block assigned to the SM, such as CUDA cores for normal arithmetic operations, special function units (SFUs) for transcendental arithmetic operations, and load and store units (LD/ST) for transferring data from/to cache or memory.

\begin{figure*}
\setlength{\abovecaptionskip}{-0.02cm}
\centering
\subfigure[Default sequential execution]{
\label{fig:diagram1} %% label for first
\includegraphics[width=0.295\textwidth]{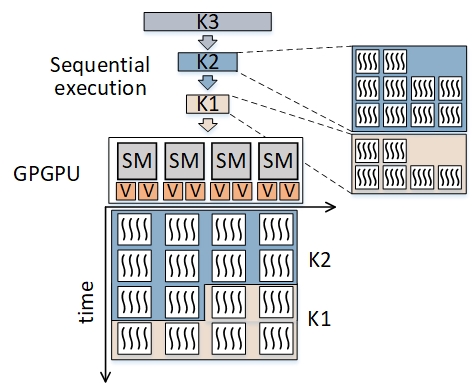}}
\hspace{0.6cm}
\subfigure[Kernel-granularity scheduling]{
\label{fig:diagram2} %% label for second subfigure
\includegraphics[width=0.295\textwidth]{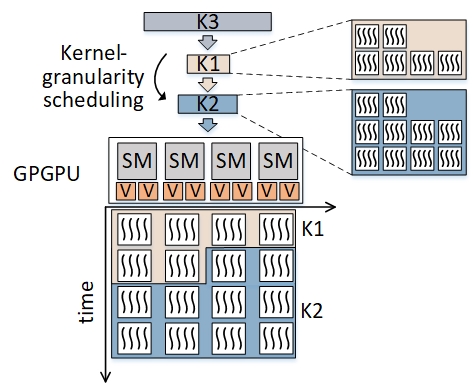}}
\hspace{0.6cm}
\subfigure[SM-granularity scheduling]{
\label{fig:diagram3} %% label for second subfigure
\includegraphics[width=0.295\textwidth]{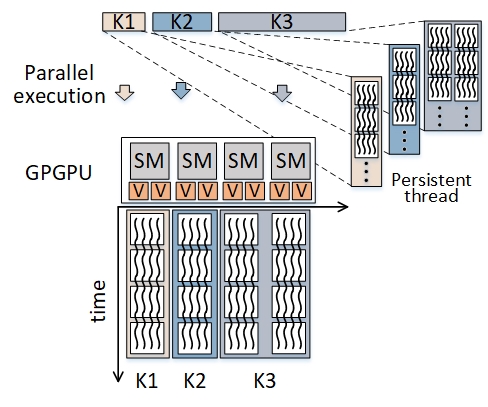}}
\caption{Comparison of three different GPU application scheduling approaches.}
\label{fig:diagram} %% label for entire figure
\vspace{-4mm}
\end{figure*}
When GPU-accelerated tasks are executed concurrently, kernels from different tasks are issued to a GPU simultaneously. When kernels are launched, the thread blocks are dispatched to all the SMs on a first-come, first-served basis. An \emph{occupancy factor},  defined as the ratio of active warps (a group of adjacent threads) on an SM to the maximum number of active warps supported by the SM, is used to describe the capacity of SMs. If the first-launched kernel is large and occupies all the GPU resources (the occupancy factor is 1), the next kernel begins its execution only when the first kernel is about to finish and resources within SMs are freed (occupancy factor below 1). To better manage GPU resources and support multiple kernels concurrently, Multi Process Service (MPS) and Multi-Instance GPU (MIG) have been introduced by NVIDIA. For example, the CUDA contexts belonging to MPS clients funnel their work through the MPS server. It allows client CUDA contexts to bypass hardware limitations associated with time sliced scheduling, and permit CUDA kernels to execute simultaneously \cite{MPS}.
\vspace{-4mm}

\subsection{Persistent Threads}
\label{sec:persistentthreads}

An off-the-shelf GPU supports only kernel-granularity scheduling, as shown in Fig. \ref{fig:diagram1}. {When kernels are launched in the GPU, if the kernel is large enough to fully occupy all the compute resources (SMs and CUDA cores) on the GPU, a GPU is only able to execute one kernel at a time by default even with Multi-Process Service (MPS).} The kernel execution orders from different tasks can be changed in kernel-granularity scheduling, as shown in Fig. \ref{fig:diagram2}.  

The persistent threads approach is a new software workload assignment solution proposed to implement finer and more flexible SM-granularity GPU partitioning \cite{gupta2012study,yu2018smguard,wu2015enabling}. Specifically, each persistent threads block links multiple thread blocks of
one kernel and is assigned to one SM to execute for the entire hardware execution lifetime of the kernel. For example, in Fig. \ref{fig:diagram3}, the first thread block in kernel 1 (K1) links the other thread blocks in K1 to form a big linked thread block. When this first thread block is executed by one SM, the other thread blocks in K1, which are linked by the first block, will also be executed in the first SM. Thus, K1 takes one SM to execute. Similarly, in kernel 3 (K3), the first two thread blocks link the other thread blocks and form two big linked thread locks. Thus, the kernel 3 (K3) takes two SMs to execute.
When the numbers of linked thread blocks are changed, the resulting number of persistent threads blocks controls how many SMs (i.e., GPU resources) are used by a kernel. In addition, when there are remaining available SMs, CUDA introduces CUDA Streams that support concurrent execution of multiple kernels. By exploiting persistent threads and CUDA Streams, we can explicitly control the number of SMs used by each kernel and execute kernels of different tasks concurrently to achieve SM-granularity scheduling. Persistent threads enabled SM-granularity scheduling fundamentally improves schedulability of parallel GPU applications by exploiting finer-grained parallelism. \vspace{-2mm}

\subsection{Multi-Segment Self-Suspension Model}
\label{sec:suspension}
In the multi-segment self-suspension model, a task $\tau_i$ has ${m_{i}}$ execution segments and ${m_{i}-1}$ suspension segments between the execution segments. So task $\tau_i$ with deadline $D_i$ and period $T_i$ is expressed as a 3-tuple:
\vspace{-2mm}
\begin{equation*}
\tau_i = \big( (L_{i}^{0}, S_{i}^{0}, L_{i}^{1}, ..., S_{i}^{m_{i}-2}, L_{i}^{m_{i}-1}), D_{i}, T_{i} \big)
\label{eq:self-suspension model}
\vspace{-2mm}
\end{equation*}
where $L_{i}^{j}$ and $S_{i}^{j}$ are the lengths of the $j$-th execution and suspension segments, respectively. $[\Slo_{i}^{j}, \Sup_{i}^{j}]$ gives the lower and upper bounds of the suspension length $S_{i}^{j}$.  $\Lup_{i}^{j}$ is the upper bound on the length of the execution segment $L_{i}^{j}$.
The analysis in \cite{schonberger2018schedulability} bounds the worst-case response time of a task under the multi-segment self-suspension model, which is summarized below.
\vspace{-2mm}
\begin{lemma}
\label{lem:W}
The following workload function $W_{i}^{h}(t)$ bounds on the maximum amount of execution that task $\tau_i$ can perform during an interval with a duration $t$ and a starting segment $L_{i}^{h}$:
\vspace{-3mm}
%of the $hth$ computation segment in task $i$ in an interval length $t$ can be expressed as:
\begin{equation*}
\begin{aligned}
%\label{eq:W}
& W_{i}^{h}(t) = \sum_{j=h}^{l} \Lup_{i}^{j \bmod m_{i}}+ \\
&\quad\quad \min \Big(\Lup_{i}^{(l+1) \bmod  m_{i}}, t-\sum_{j=h}^{l} \big(\Lup_{i}^{j \bmod  m_{i}}+S_{i}(j) \big) \Big)
\end{aligned}
\end{equation*}
where $l$ is the maximum integer satisfying the following condition:
\vspace{-3mm}
\begin{equation*}
\begin{aligned}
\sum_{j=h}^{l} \big(\Lup_{i}^{j \bmod  m_{i}}+S_{i}(j) \big) \leq t
\end{aligned}
%\label{eq:W2}
\vspace{-2mm}
\end{equation*}
and $S_{i}(j)$ is the minimum interval-arrival time between execution segments $L_i^j$ and $L_i^{j+1}$, which is defined by:
\begin{equation*}
S_{i}(j) = \left\{
             \begin{array}{@{}l@{\mkern-36mu}r@{}}
\displaystyle {\Slo}_{i}^{j \bmod m_{i}} & \mbox{if\ } j \bmod m_{i} \neq (m_{i}-1)  \\
\displaystyle T_{i}-D_{i} & \mbox{else if\ }  j = m_{i}-1  \\ %j \leq m_{i}
\displaystyle T_{i}- \sum_{j=0}^{m_i-1}\Lup_{i}^j -\sum_{j=0}^{m_i-2}\Slo_{i}^j &  \mbox{otherwise} 
             \end{array}
\right.
\end{equation*}
\vspace{-5mm}
\end{lemma}

%The maximum workload function $W_{i}(t)$ of task $i$ in an interval length $t$ is:
%\begin{equation}
%W_{i}(t) = max_{h_{i}\in[0,m_{i}-1]}{W_{i}^{h_{i}}(t)}
%\label{eq:workload function}
%\end{equation}

Then the response time of execution segment $L_i^j$ in task $\tau_k$ can be bounded by calculating the interference caused by the workload of the set of higher-priority tasks $hp(k)$.  
\vspace{-2mm}
\begin{lemma}
\label{lem:Rkj}
The worst-case response time $\Rup_{k}^{j}$ is the smallest value that satisfies the following recurrence:
\begin{equation*}
\Rup_{k}^{j} = \Lup_{k}^{j} + \sum_{\tau_{i}\in hp(k)}  \max_{h\in[0,m_{i}-1]}{W_{i}^{h}}  (\Rup_{k}^{j})
%\label{eq:Rkj}
\end{equation*}
\vspace{-4mm}
%where $W_{i}(t)$ is the maximum workload function and $hp(k)$ denotes the set of the tasks with higher priority than task $k$.
\end{lemma}

Hence, the response time of task $\tau_k$ can be bounded by either taking the summation of the response times of every execution segments and the total worst-case suspension time, or calculating the total interference caused by the workload of the set of higher-priority tasks $hp(k)$ plus the total worst-case execution and suspension time.
\vspace{-2mm}
\begin{lemma}
\label{lem:Rk}
Hence, the worst-case response time $\Rup_{k}$ of task $\tau_k$ is upper bounded by the minimum of $\RupA_{k}$ and $\RupB_{k}$, where:
\vspace{-2mm}
\begin{equation}
\RupA_{k} = \sum_{j=0}^{m_k-2}\Sup_{k}^j + \sum_{j=0}^{m_{k}-1} \Rup_{k}^{j}
\label{eq:Rk1}
\end{equation}
\vspace{-2mm}
and $R2_{k}$ is the smallest value that satisfies the recurrence:
\begin{equation}
\RupB_{k} = \sum_{j=0}^{m_k-2}\Sup_{k}^j + \sum_{j=0}^{m_k-1}\Lup_{k}^j + \sum_{\tau_{i}\in hp(k)} \max_{h\in[0,m_{i}-1]}{W_{i}^{h}} (\RupB_{k})
\label{eq:Rk2}
\end{equation}
\end{lemma}

\vspace{-2mm}

\section{CPU and Memory Model}
\label{sec:CPU_memory}
{In this work, we target CPU-GPU heterogeneous computing systems. The heterogeneous systems may have different typologies such as different numbers of CPU cores, memory copy engines, and GPUs (and also SMs in one GPU). To propose a general system model, we start from the most fundamental case which only has one CPU core, one memory copy engine, and multiple SMs in one GPU. All applications are written as threads of a single process. This fundamental case is the minimal heterogeneous system with a full heterogeneous computing function. Any heterogeneous system can be regarded as a combination of this fundamental case. This work aims to study the fundamental CPU-GPU heterogeneous computing system real-time scheduling problem as the first step and then extend the study to multiple GPUs in the future.}
\vspace{-2mm}

\subsection{CPU Modeling}
As represented in Fig. \ref{fig:diagram0.jpg}, a complete GPU application has multiple segments of CPU code, memory copies between the CPU and GPU, and GPU code (also called GPU kernels). The CPU executes serial instructions, e.g., for communication with IO devices (sensors and actuators) and launches memory copies and GPU kernels. When a CPU executes serial instructions, it naturally behaves as a single-threaded application without parallelism. When the CPU code launches memory copies or GPU kernels, these instructions will be added into multiple FIFO buffers called a ”CUDA stream”. The memory copies and GPU kernels, which are in different CUDA streams, can execute in parallel if there are remaining available resources. The execution order of memory copies and GPU kernels in a single CUDA stream can be controlled by the order in which they are added to it by the CPU code. After the CPU has launched memory copies and GPU kernels into a CUDA stream, it will immediately execute the next instruction, unless extra synchronization is used in the CPU code to wait for the memory copies or GPU kernels to finish. Thus, the CPU segments can be modeled as serial instructions in one thread.
\vspace{-2mm}

\subsection{Memory Modeling}
Memory copying between the CPU and GPU execution units includes two stages. In the first stage which is also called global memory copy, data is copied between the CPU memory and the GPU memory through a single {peripheral component interconnect express} (PCIe) or through a {network on chip} (NoC). The PCIe and NoC offer packet-based and full-duplex communication between any two endpoints. The number of global memory copies that can happen simutanously are determined by the number of copy engines provided by GPUs. For example, GeForce GTX TITAN Black and Jetson TX2 have 1 copy engine; 1080TI, TITAN X and NVIDIA Xavier \cite{yang2018avoiding} have 2 copy engines. In this work, we assume that there is only one copy engine in the minimal heterogeneous system model, which has one CPU core and multiple heterogeneous cores. Also, the memory copy through PCIe/NoC is non-preemptive once it starts.
The GPU and other accelerators mainly provide two types of first stage memory movement \cite{amert2017gpu, otterness2017evaluation}: direct memory copy (also called traditional memory) and unified memory (introduced in CUDA 6.0). Direct memory copy uses traditional memory, where data must be explicitly copied from CPU to GPU portions of DRAM. Unified memory is developed from zero-copy memory where the CPU and the GPU can access the same memory area by using the same memory addresses between the CPU and GPU. In the following discussion, we focus mainly on direct memory copy, but our approach can also be directly applied to unified memory by setting explicit copy length to zero.
The second stage is the {memory access} from the GPU’s execution units to the GPU cache or memory. The GPU adopts a hierarchical memory architecture. Each GPU SM has a local L1 cache, and all SMs share a global L2 cache and DRAM banks. These memory accesses happen simultaneously with the kernel’s execution. Therefore, the second stage memory operation is measured and modeled as part of the kernel execution model. Although run-time memory factors, such as the state of the row buffers in the first stage and contention on GPU memory or cache in the second stage, would impact memory copy time, we have to simplify the memory model with static factors given the consideration of real-time scheduling complexity. Therefore, we assume that the memory copy time between CPU memory and GPU memory is a linear function of the copied memory size.
\vspace{-2mm}

\section{Modeling and Management of GPU Fine-Grain Partitioning}
\label{sec:GPU_model}

%\vspace{-1mm}
Following the persistent thread technique, this section introduces the modeling and management for GPU fine-grain partitioning. The proposed technique in this section takes both throughput and flexibility into account. It develops the system foundation for GPU real-time scheduling with high schedulability.
\vspace{-2mm}

\subsection{Kernel Execution Model}
\label{kernelmodel}
\begin{figure}
\centering
\setlength{\abovecaptionskip}{-0.05cm}
\hspace{-4mm}
\subfigure[computation]{
\label{fig:synthetic_kernel1.jpg} %% label for second subfigure
\includegraphics[width=0.122\textwidth]{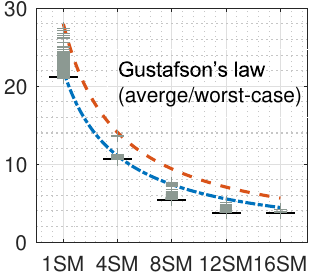}}
\hspace{-2mm}
\subfigure[memory]{
\label{fig:synthetic_kernel2.jpg} %% label for second subfigure
\includegraphics[width=0.122\textwidth]{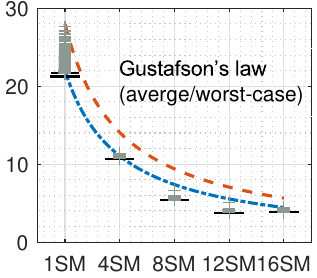}}
\hspace{-2mm}
\subfigure[branch]{
\label{fig:synthetic_kernel2.jpg} %% label for second subfigure
\includegraphics[width=0.122\textwidth]{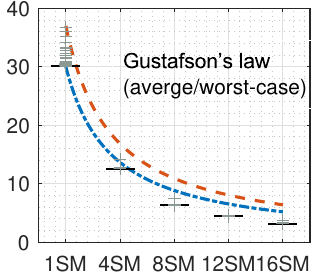}}
\hspace{-2mm}
\subfigure[special]{
\label{fig:synthetic_kernel2.jpg} %% label for second subfigure
\includegraphics[width=0.122\textwidth]{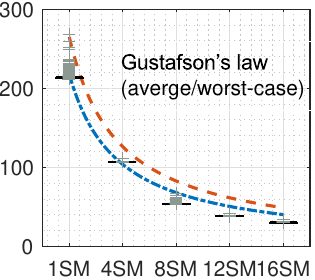}}
\hspace{-4mm}
\caption{{Kernel execution time trends.}}
\label{fig:synthetic} %% label for entire figure
\vspace{-5mm}
\end{figure}

To understand the relationship between the execution time of a kernel and the number of SMs assigned via persistent threads, we conducted the following experiments. We use five synthetic kernel benchmarks that utilize different GPU resources: a computation kernel, consisting mainly of arithmetic operations; a branch kernel containing large number of conditional branch operations; a memory kernel full of memory and register visits; a special-function kernel with special mathematical functions, such as sine and cosine operations; and a comprehensive kernel including all these arithmetic, branch, memory, and special mathematical operations. Each kernel performs 1000 floating-point operations on a $2^{15}$-long vector.

We first run each kernel separately with a fixed workload for 1000 times and record its corresponding execution time with increasing numbers of assigned SMs, as shown in Fig.~\ref{fig:synthetic_kernel1.jpg}. Next, we examine the kernel execution time with increasing kernel sizes and different numbers of assigned SMs. Fig.~\ref{fig:synthetic_kernel2.jpg} shows that the comprehensive kernel and the other types of kernels have similar trends. {From the boxplot, we can see that the kernel execution time follows the formula of Gustafson’s law \cite{gustafson1988reevaluating} and can be expected of a system whose resources are more flexible:}
\begin{equation}
    S = N + (1-N)s
\label{eq:1}
\end{equation}
Where $N$ is the number of assigned SMs, $s$ is the serial fraction of the workload (which does not benefit from parallelism), and $S$ is the estimated speedup. The S speedup in latency is normalized to the kernel only with computation instruction.
From the architecture perspective, the GPU kernels are fully parallel workloads, which can utilize all allocated SMs. The only sequential execution is when the GPU is copying data and launching the kernel. We can also observe that the execution time of a GPU kernel has low variation because it benefits from a single-instruction multiple-threads (SIMT) architecture, in which single-instruction, multiple-data (SIMD) processing is combined with multithreading for better parallelism.
\vspace{-4mm}

\subsection{Interleaved Execution and Virtual SM}
\begin{figure}
\setlength{\abovecaptionskip}{-0.02cm}
\centering
\includegraphics[width=0.5\textwidth]{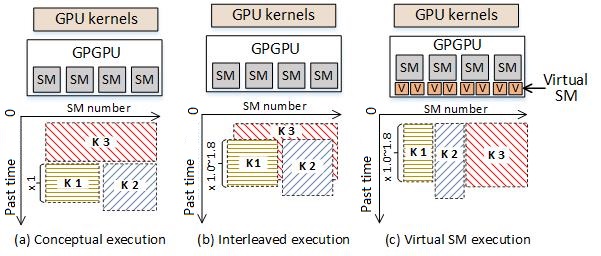}
\caption{Virtual SM model for interleaved execution}
\label{fig:interleaved_execution}
\vspace{-6mm}
\end{figure}

Through a close comparison of the GPU kernel execution and the design of GPU architectures, we find that the system throughput can be further improved by exploiting interleaved execution of GPU kernels. On a GPU with $M$ SMs, naive SM-granularity scheduling can first concurrently execute the $K_1$ and $K_2$ kernels, each with $M/2$ persistent threads blocks, and then execute the K3 kernel with $M$ persistent threads blocks, as shown in Fig.~\ref{fig:interleaved_execution}(a). Each block requires one SM to execute one persistent thread at a time.

On the other hand, an SM actually allows the parallel execution of two or more persistent threads blocks to overlap if the current SM occupancy factor is below 1 which means the number of active warps is less than the maximum \cite{NVIDIA_Occupancy, kim2016automatically}.
This interleaved execution is similar to the hyper-threading in conventional multithreaded CPU systems that aims to improve computation performance. For example, in an NVIDIA GTX 1080 TI, one SM can hold 2048 software threads, whereas one thread block can have at most 1024 software threads. Thus, two or more thread blocks can be interleaved and executed on one SM. One important consequence of interleaved execution is that the execution time of a kernel increases. Therefore, to improve GPU utilization and efficiency, we can launch all three kernels, as illustrated in Fig.~\ref{fig:interleaved_execution}(b), where kernel 1 and kernel 2 will simultaneously execute with kernel 3. The execution latency of each kernel is increased by a factor called the interleaved factor, which ranges from 1.0 to 1.8 in the following experiments.

\begin{figure}
\setlength{\abovecaptionskip}{-0.02cm}
\centering
\subfigure[On computation kernel]{
\label{fig:memory_time_distribution_100K} %% label for second subfigure
\includegraphics[width=0.2\textwidth]{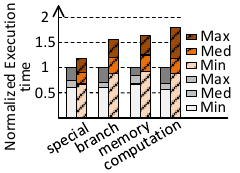}}
\subfigure[On memory kernel]{
\label{fig:memory_time_distribution_1M} %% label for second subfigure
\includegraphics[width=0.2\textwidth]{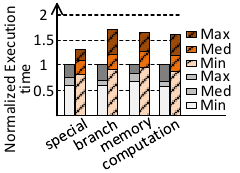}}
\subfigure[On branch kernel]{
\label{fig:memory_time_distribution_10M} %% label for second subfigure
\includegraphics[width=0.2\textwidth]{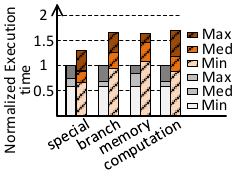}}
\subfigure[On special kernel]{
\label{fig:memory_time_distribution_100M} %% label for second subfigure
\includegraphics[width=0.2\textwidth]{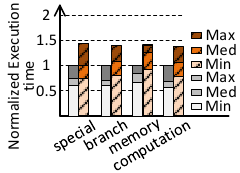}}
\caption{Characterization of the latency extension ratios of interleaved execution}
\label{fig:interleaved_execution_2} %% label for entire figure
\vspace{-4mm}
\end{figure}

We propose a virtual SM model to capture this interleaved execution of multiple GPU kernels, as shown in Fig.~\ref{fig:interleaved_execution}(c). In particular, we double the number of physical SMs to get the number of virtual SMs. Compared with a physical SM, a virtual SM has a reduced computational ability and hence a prolonged execution time, the length of which is related to the type of instructions in the interleaved kernel. To understand the interleaved ratio, 
we empirically measured the execution time of a synthetic benchmark when it was interleaved with another benchmark. Fig.~\ref{fig:interleaved_execution_2} illustrates the minimum, median, and maximum interleaved execution time, colored from light to dark, normalized over the worst-case execution time of the kernel without interleaving, where the left bar is without interleaving and right bar is with interleaving. We can see that the interleaved execution ratio is at most $1.45\times$, $1.7\times$, $1.7\times$, and $1.8\times$ for special, branch, memory and computation kernels, respectively. The proposed virtual SM model improves throughput by $11\%\sim38\%$ compared to the naive non-interleaved physical SM model. The number of virtual SMs is determined by how many threads can be physically simultaneously executed on one SM. {In this work, we use the NVIDIA GTX 1080 TI GPU as an example. In this GPU one physical SM can hold and execute 2048 threads and one thread block has 1024 threads at most. If one thread block is at its highest capacity with 1024 threads, two thread blocks are executed on one physical SM. Therefore, one physical SM will be mapped to two virtual SMs. If one thread block only has 512 threads, four thread blocks are executed on one physical SM. and one physical SM will be mapped to four virtual SMs. Therefore, by limiting the maximum number of threads in one thread block (in writing the GPU kernels) to $2048/V_{SM}$, we can generate $V_{SM}$ virtual SMs. \textcolor{black}{In this work, we follow the default setting of the NVIDIA GTX 1080 TI GPU where the maximum number of threads in one thread block is 1024.}} \vspace{-2mm}

\subsection{Workload Pinning and Self-Interleaving}
Using the persistent threads and interleaved execution techniques, multiple tasks can be executed in parallel, and the interleaved execution further improves GPU performance. In real GPU systems, such as NVIDIA GPUs, a hardware scheduler is implemented that allocates the thread blocks to SMs in a greedy-then-oldest manner \cite{gpgpusim}. Thus, at run time, the thread blocks from a kernel are interleaved and executed with thread blocks from other possible kernels, and the interleaved execution ratio is different when different kernels are interleaved and executed, as shown in Fig. \ref{fig:interleaved_execution_2}. To guarantee a hard deadline, each kernel has to adopt the largest interleaved execution ratio when this kernel is interleaved and executed with other possible kernels. However, using the highest interleaved execution ratio cannot avoid underestimation of the GPU computation ability. Therefore, we introduce workload pinning which pins the persistent threads blocks to specific SMs, and {self-interleaving} where the kernel interleaves with itself on its pinned SMs.

\begin{algorithm}
\caption{Pseudo Code of Pinned Self-Interleaving Persistent Threads Pseudo Code}
{
\begin{flushleft}
\textit{\small{// Get the ID of current SM with assemble language}}\\
\textbf{static \_\_device\_\_} \_\_inline \textbf{\_\_ uint32\_t} \_\_mysmid()\quad\{ \\   
  \textbf{uint32\_t} smid;    \\
  \textbf{asm volatile} ("mov.u32 \%0, \%\%smid;" : "=r"(smid));   \\ 
  \textbf{return} smid;\quad\}\\
  \quad\\
\textit{\small{// Kernel pinned to desired\_SMs with self-interleaved persistent threads}}\\
\textbf{\_\_global\_\_ void} kernel (\textbf{int} *desired\_SMs, ...)\{ \\
\textbf{int} SMs\_num = length(desired\_SMs); \\
\textbf{int} Real\_SM\_id; \\
Real\_SM\_id = \_\_mysmid(); \textit{\small{// Get the ID of current SM}} \\
\textit{\small{//Excute on desired SMs, otherwise return}}\\
\textbf{if}(Real\_SM\_id == $\forall$k, desired\_SMs[k]) \{ \\
\quad \textit{\small{//Get the global thread index: tid}}\\
\quad \textbf{int} tid = threadIdx.x + k * blockDim.x;\\
\quad \textit{\small{//off\_set links to the next thread block by persistent threads}}\\
\quad \textbf{int} off\_set = blockDim.x * SMs\_num;\\
\textit{\small{//Divide N threads inside a kernel to V pieces [0 N/V) and [N/V 2N/V) ... and [(V-1)N/V N) from same kernel interleaved execute with each other. From the kernel perspective, the kernel interleaved execute with itself.}}\\
\quad \textbf{if}(blockIdx.x $<$ virtual\_SM/V) \{\\
\quad\quad\textbf{for}(\textbf{int} i = tid; i $<$ N/V; i += off\_set) \{\\
\quad\quad\quad Execute on thread i;\}\}\\ 
\quad \textbf{else \!if}(virtual\_SM\!/V\!\! $\leq$ \!\! blockIdx.x \!\!$<$\!\! 2virtual\_SM\!/V\!) \!\!\{ \\
\quad\quad\textbf{for}(\textbf{int} i = tid + N/V; i $<$ 2N/V; i += off\_set) \{\\
\quad\quad\quad Execute on thread i;\}\}\\
\quad \textbf{else \!if}(2virtual\_SM\!/\!V\!\!\! $\leq$ \!\!\! blockIdx.x\!\! $<$ \!\!3virtual\_SM\!/\!V\!) \!\!\{ \\
\quad\quad... \}\\
\}\\
\textbf{return}; \}\\
\quad\\
\textit{// Kernel launch}\\
\textbf{int} main () \{ \\
\textbf{int} desired\_SMs[] = \{1, 2, 4\}; \textit{//The desired SM\_id in this example is 1, 2, 4} \\
\textbf{dim3} gridsize (number of virtual SM);\\
\textbf{dim3} blocksize (Max number of threads per block);\\
kernel $<<<$ gridsize, blocksize, ..., stream $>>>$ (desired\_SMs, ...);\\
\textbf{return 0}; \}
\end{flushleft}
\label{Algorithm: persistent_thread}
}
\vspace{-2mm}
\end{algorithm}

Workload pinning is implemented by launching Vm (m is the number of physically assigned SMs) persistent threads blocks in each kernel, which is also the number of virtual SMs, so that all virtual SMs will finally have one persistent threads block to execute. {If the SM is the targeted pinning SM, the thread block will begin to execute. Persistent threads blocks assigned to undesired SMs (untargeted pinning SMs), will simply return, which takes about 10s µs. When a persistent threads block is assigned to the correct SM, it will not only execute its own workload, but will also execute the workloads from blocks assigned to the undesired SMs. Thus, the kernel is actually executed on the desired SMs, and the undesired SMs execute an empty block within a negligible time. Therefore, the overhead from the GPU fine-grain partitioning is around 10s us.} 

{A persistent threads with pinned self-interleaving design and implementation is implemented in the CUDA code, which is detailed described in Algorithm \ref{Algorithm: persistent_thread}. For the usage of the proposed GPU fine-grain partitioning, the researchers will follow the persistent threads style and add the comparison of SM id number at the beginning of the GPU kernel to realize workload pinning and self-interleaving.}
\vspace{-2mm}

%To tolerate the interference, such as run-time memory factors in memory copy and the dynamics in CPU/GPU executions, the model parameters of the above CPU, memory copy and GPU kernels are noted with best-case and worst-case execution time.
\section{Practical Full System Scheduling}
\label{sec:schedule}

In this section, we first introduce the RT-GPU scheduling algorithm, and then develop the corresponding timing analysis. 
One of the key challenges of deriving the end-to-end response times is to simultaneously bound the interference on CPU, GPU, and memory bus.
{As the start of memory copy and dispatch of GPU kernel are initialized by the CPU, the scheduling of the full system is managed by the scheduler on the CPU side, i.e. the real-time scheduler in Linux. Therefore, the start of memory copy and GPU kernel will immediately follow its previous CPU segment and memory copy.}

\vspace{-4mm}

\begin{figure}
\centering
\includegraphics[width=0.48\textwidth]{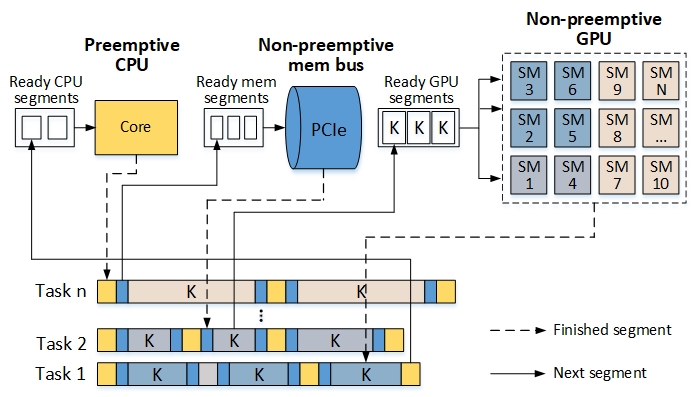}
\caption{GPU tasks real-time scheduling model.}
\label{fig:scheduling_model}
\vspace{-5mm}
\end{figure}

%\vspace{-4mm}
\subsection{Task Model}
\label{sec:task_model}
%As discussed in Section \ref{sec:CPU_memory} and \ref{sec:GPU_model}, each GPU task has CPU segments, memory copy segments and GPU kernel segments, which must be executed in order. The CPU code executes logic operations serially and usually does not contain parallel execution inside one task. The memory movement from CPU (host) and GPU (device) is through a PCIe or network-on-chip interface in an explicit or implicit manner. Helped by the persistent technique with workload pinning, now the many-SM GPU can be used as a classic manycore system and the GPU kernel can be executed fully in parallel on specific SMs but the execution is in a non-preemptive manner \cite{GPUpreem}. 

%Based on previous discussion and system work, GPU tasks scheduling model is shown in Fig. \ref{fig:scheduling_model}. In this scheduling model, there is a preemptive CPU core executing the CPU segments, a non-preemptive communication interface (bus) for memory copy segments, a GPU with many cores (SMs) for the GPU kernels. This model is not limited to general commercial embedded, desktop and server GPU systems, it can also be appropriate for the general heterogeneous computing system such as CPU+FPGA, CPU+accelerator architectures and so on.

%Since the GPU tasks are embarrassingly parallel workloads and there is no synchronization between tasks

Leveraging the platform implementation and the CPU, memory and GPU models discussed in previous sections, the model for the parallel real-time tasks executing on a CPU-GPU platform is shown in Fig. \ref{fig:scheduling_model}. We consider a task set $\tau$ comprised of $n$ parallel tasks, where $\tau = \{ \tau_1, \tau_2, \cdots, \tau_n \}$. Each task $\tau_i$, where $1\le i\le n$, has a relative deadline $D_{i}$ and a period (minimum inter-arrival time) $T_{i}$. In this work, we restrict our attention to constrained-deadline tasks, where $D_i \leq T_i$, and tasks with fixed task-level priorities, where each task is associated with a unique priority. More precisely, when making scheduling decisions on any resource, such as CPU and bus, the system always selects the segment with the highest priority among all available segments for that resource to execute. Of course, a segment of a task only becomes available if all the previous segments of that task have been completed.

On a CPU-GPU platform, task $\tau_i$ consists of $m_i$ CPU segments, $2m_i-2$ memory copy segments, and $m_i-1$ GPU segments. 
As discussed in Section~\ref{kernelmodel}, a GPU segment $G_{i}^{j}$ models the execution of a GPU kernel on interleaved SMs using total work $\GW_{i}^{j}$, critical-path overhead $\GL_{i}^{j}$, and interleaved execution ratio $\alpha_{i}^{j}$, i.e., $G_{i}^{j} = (\GW_{i}^{j}, \GL_{i}^{j}, \alpha_{i}^{j} )$.
Thus, task $\tau_i$ can be characterized by the following tuple:
\vspace{-2mm}
\begin{equation}
\begin{split}
\tau_i = 
\Big( & \big(\CL_{i}^{0}, \ML_{i}^{0}, G_{i}^{0}, \ML_{i}^{1}, 
 \CL_{i}^{1}, \ML_{i}^{2}, G_{i}^{1}, \ML_{i}^{3}, \\ 
& \cdots, \CL_{i}^{j}, \ML_{i}^{2j}, G_{i}^{j}, \ML_{i}^{2j+1}, \cdots, 
 \CL_{i}^{m_{i}-2}, \\ & \ML_{i}^{2m_{i}-4}, G_{i}^{m_{i}-2}, \ML_{i}^{2m_{i}-3}, 
 \CL_{i}^{m_{i}-1} \big), D_{i}, T_{i} \Big)
\label{eq:intrinsic_Switched_capacitor_loss}
\end{split}
\vspace{-5mm}
\end{equation}
where $\CL_{i}^{j}$ and $\ML_{i}^{j}$ are the execution times of the \mbox{$(j+1)$-th} CPU and memory copy segments, respectively. In addition, we use $\widecheck{\ \ }$ and $\widehat{\ \ }$ to denote the lower and upper bound on a random variable. For example, $\CLup_{i}^{j}$ and $\CLlo_{i}^{j}$ are the upper and lower bounds on execution times of the $(j+1)$-th CPU segment of $\tau_i$, respectively.

%If ${m_{i}}$ is 1, there is only one CPU segment of $\tau_i$, and the task $\tau_i$ does not have memory copy or GPU codes. This model is a basic configuration where there is one CPU execution, one time memory movement and one GPU execution in one CPU, memory. and GPU data flow. Any task configurations can be generated from this configuration by setting the length of undesired segments to zero. A task system $\tau$ is said to be an implicit-deadline system if its deadline is equal to its period and a constrained-deadline system if the relative deadline of each task is no larger than its period.

To derive the end-to-end response time $R_i$ of task $\tau_i$, we will analyze the response times $GR_{i}^{j}$, $MR_{i}^{j}$, and $CR_{i}^{j}$ of each individual GPU, memory copy, and CPU segments, respectively, and calculate their lower and upper bounds in the following subsections. \vspace{-2mm} 
%Additionally, since the response times of one type of segments will be viewed as suspension when analyzing the other type of segments, both the lower and upper bounds on the response times will be derived.

%$[{\mathop{CR_{i}^{j}}\limits^{\vee}_{}} \; {\mathop{CR_{i}^{j}}\limits^{\wedge}_{}}]$, 
%$[{\mathop{MR_{i}^{j}}\limits^{\vee}_{}} \; {\mathop{MR_{i}^{j}}\limits^{\wedge}_{}}]$, and 
%$[{\mathop{GR_{i}^{j}}\limits^{\vee}_{}} \; {\mathop{GR_{i}^{j}}\limits^{\wedge}_{}}]$,  

%$[{\CRlo_{i}^{j}} \; {\CRup_{i}^{j}}]$, 
%$[{\MRlo_{i}^{j}} \; {\MRup_{i}^{j}}]$, and 
%$[{\GRlo_{i}^{j}} \; {\GRup_{i}^{j}}]$, 

%\vspace{-6mm}
\subsection{Federated Scheduling for GPU Segments}
\label{sec:GR}
%In this work, we focus on the parallel real-time scheduling problem on a GPU comprised of $N_{SM}/2$ physical SMs modeled as $N_{SM}$ virtual SMs and a taskset with $n$ tasks.
%We consider the sporadic real-time tasks, where the $jth$ GPU segment in task $i$ has an {implicit deadline} $GD_i^{j}$, which is equal to its period $GT_i^{j}$. 

%Since the GPU kernels are embarrassingly parallel workloads, as discussed in Section \ref{sec:GPU_model}, a GPU segment in task $\tau_i$ where $1\le i\le n$ with multiple kernels can be modeled similarly using the classic {parallel synchronous task model}. We model the {total work} $\GW_{i}^{j}$ of a kernel $G_{i}^{j}$ using the total execution time of the kernel on a single SM in isolation, i.e., without any interleaving. The overhead of launching kernel is denoted as {critical-path overhead} $\GL_{i}^{j}$.
%Therefore, the total work $\GW_i^{j}$ of the GPU kernels in task $\tau_i$ is $\GW_i^{j}$ and the critical-path overhead is $\GL_i^{j}$. The {utilization} $u_i^{j}$ of this GPU segment is defined as $u_i = \GW_i^{j}/T_i^{j}$. To capture a GPU segment's execution on self-interleaved SMs, we define the total execution time of this GPU segment on a single SM with two virtual SMs interleaved with itself as the {total virtual work} $\GW_i^{j} \alpha_i^j$, where the $\alpha_i^j$ is defined as the {interleaved execution ratio} of this segment.
For executing the GPU segments of the $n$ tasks on the shared GPU with ${\VGN}$ virtual SMs (i.e., ${\GN}$ physical SMs), we propose to generalize {federated scheduling}, a scheduling paradigm for parallel real-time tasks on general-purpose multi-core CPUs, to scheduling parallel GPU segments. The key insight of federated scheduling is to calculate and assign the minimum number of dedicated resources needed for each parallel task to meet its deadline.

Specifically, we allocate $\VGN_i$ dedicated virtual SMs to each task $\tau_i$, such that its GPU segment $G_i^j$ can start executing immediately after the completion of the corresponding memory copy $\ML_i^{2j}$. In this way, the mapping and execution of GPU kernels to SMs are explicitly controlled via the persistent thread and workload pinning interfaces, so the effects caused by the black-box internal scheduler of a GPU are minimized. Additionally, tasks do not need to compete for SMs, so there is no blocking time on the non-preemptive SMs. Furthermore, via the self-interleaving technique, we enforce that different GPU kernels do not share any physical SMs. Therefore, the interference between different GPU segments is minimized, and the execution times of GPU segments are more predictable.

In summary, each task $\tau_i$ is assigned with $\VGN_i$ dedicated virtual SMs where each of its GPU segments self-interleaves and has an interleaved execution ratio $\alpha_i^j$. In Section~\ref{sec:algo}, we will present the algorithm that determines the SM allocation to tasks. Here, for a given allocation, we can easily extend the formula in Section~\ref{kernelmodel} to obtain the following lemma for calculating the response time $\GR_i^j$ of a GPU segment $G_i^j$.

%\begin{lemma}
%\label{lem:ncores}
%\textbf{(From \cite{FS})}
%If the GPU segment has a total virtual work $GW'_i^{j}$ and a critical-path overhead of $GLL_i^{j}$, then it is schedulable and can meet its implicit deadline $GD_i^{j}$ with $GN_i$ dedicated virtual SMs, where $GN_i \geq \frac{GW'_i^{j}-GLL_i^{j}}{GD_i^{j}-GLL_i^{j}} $.
%\end{lemma}
\begin{lemma}
\label{lem:ncores}
If the GPU segment $G_i^j$ has a total work in range $[\GWlo_i^{j}, \GWup_i^{j}]$, a critical-path overhead in range $[0, \GLup_i^{j}]$ and an interleaved execution ratio in range $[1, \alpha_i^j]$, then when running on $\VGN_i$ dedicated virtual SMs, its response time is in $[\GRlo_i^{j}, \GRup_i^{j}]$ where
\[
\GRlo_i^j = \frac{\GWlo_i^{j}}{\VGN_i}, \mbox{\ and \ }
\GRup_i^j = \frac{\GWup_i^{j}\alpha_i^j - \GLup_i^{j}}{\VGN_i}+\GLup_i^{j}.
\]
\end{lemma}
\begin{proof}
{
The lower bound $\GRlo_{i}^{j}$ is the minimum execution time of this GPU segment on $VGN_i$ virtual SMs. In the best case, there is no critical-path overhead and no execution time inflation due to interleaved execution. The minimum total virtual work $\GWlo_i^{j}$ is executed in full parallelism on $VGN_i$ virtual SMs, which gives the formula for $\GRlo_{i}^{j}$.

In the worst case, the maximum total virtual work is $\GWup_i^{j}\alpha_i^j$, as it demands the most computation and thus longest execution time. Additionally, the maximum critical-path overhead $\GLup_i^{j}$ captures the maximum overhead of launching the kernel, which run serially and cannot benefit from parallelism. Since $\GLup_i^{j}$ is a constant overhead and is not affected by self-interleaving and multiple virtual SMs, we do not need to apply the interleaved execution ratio $\alpha_i^j$ to $\GLup_i^{j}$. After deducting the critical-path overhead using to Gustafson's law in Equation~\ref{eq:1}, the remaining GPU computation is embarrassingly parallel on $VGN_i$ virtual SMs, which results the formula of $\GRup_{i}^{j}$. 
}
\end{proof}\vspace{-2mm}

Note that Lemma~\ref{lem:ncores} calculates both the lower and upper bounds on the response time of GPU segment $G_i^j$, because both bounds are needed when analyzing the total response time of task $\tau_i$. Both the lower and upper bounds can be obtained by profiling the execution time of GPU segments.

%Note that the above Lemma~\ref{lem:ncores} for calculating the minimum number of virtual SMs is similar to the Theorem 2 from previous work~\cite{FS} for calculating the minimum number of CPUs. Therefore, we propose the real-time GPU scheduling algorithm as follows.
%Given a task set $\tau$, our algorithm first uses Lemma~\ref{lem:ncores} to calculate the number of virtual SMs $m_i$ assigned to GPU segments in each task $\tau_i$. 

%then our algorithm {admits} the task set; otherwise, our algorithm declares the task set as unschedulable. Then the GPU response upper bounds $\GRup_{i}^{j}$ is the deadline $GD_{i}^{j}$ and the lower bounds $\GRlo_{i}^{j}$ is the shortest execution time of this GPU segment with $GN_i$ virtual SMs.

During runtime execution of schedulable task sets, the work in Section \ref{sec:GPU_model} will generate $\VGN_i$ persistent threads blocks for each GPU segment of task $\tau_i$ to execute on its own assigned $\VGN_i$ virtual SMs. 
{For the less powerful GPU with small numbers of SMs, we need to generate enough virtual SMs to make each task have its virtual SMs. This can be realized by limiting the maximum number of threads in a thread block as described by Section 4.2. For example, in the GTX1080TI GPU, each SM can execute 2048 threads simultaneously. Therefore, if each block has 1024 threads, two blocks are executed on one SM, which means one physical SM generates two virtual SMs. By limiting the maximum number of threads in one thread block to 2048/$V_{SM}$ , we can generate $V_{SM}$ virtual SMs. Therefore, in most cases, enough virtual SMs could be generated.}
\vspace{-5mm}

\subsection{Fixed-Priority Scheduling for memory copy Segments with Self-Suspension and Blocking}
\label{sec:MR}
From the perspective of executing memory copies over the bus, memory copy segments are \enquote{execution segments}; the time intervals where task $\tau_i$ spends on waiting for CPU and GPU to complete the corresponding computation are \enquote{suspension segments}. However, compared with the standard self-suspension model, memory copy over a bus has the following differences. (1) Because memory copy is non-preemptive, a memory copy segment of a high-priority task can be blocked by at most one memory copy segment of any lower-priority task if this lower-priority segment has already occupied the bus. (2) The length of suspension between two consecutive memory copies depends on the response time of the corresponding CPU or GPU segment. (3) The response times of CPU segments are related to the response times of memory copy segments, which will be analyzed in Section~\ref{sec:CR}. (4) Moreover, the lower bounds on the end-to-end response times of a task are related to the response times of all types of segments, which requires a holistic fixed-point calculation to be presented in Section~\ref{sec:algo}. Please note that above differences are not unique in the CPU-GPU system, they widely present in state-of-the-art heterogeneous systems.

We define the following memory copy workload function $\MW_{i}^{h}(t)$, which is similar to the workload function defined for standard self-suspension tasks in Section \ref{sec:suspension}. 
\begin{lemma}
\label{lem:MW}
$\MW_{i}^{h}(t)$ bounds the maximum amount of memory copy that task $\tau_i$ can perform during an interval with a duration $t$ and a starting memory copy segment $\ML_{i}^{h}$, where:
\begin{equation*}
\begin{aligned}
\MW_{i}^{h}(t) =& \sum_{j=h}^{l} \MLup_{i}^{j \bmod 2m_{i}-2 } + \min \Big(\MLup_{i}^{(l+1) \bmod 2m_{i}-2 },\\
&\qquad\qquad  t-\sum_{j=h}^{l} \big(\MLup_{i}^{j \bmod 2m_{i}-2 }+\MS_{i}(j) \big) \Big)
\end{aligned}
%\label{eq:MW}
\end{equation*}
where $l$ is the maximum integer satisfying the following condition:
\begin{equation*}
\begin{aligned}
%\sum_{j=h}^{l}(ML_{i}^{j \bmod  m_{i}}+S_{i}^{j}) \leq t
\sum_{j=h}^{l} \big(\MLup_{i}^{j \bmod 2m_{i}-2 }+\MS_{i}(j) \big) \leq t
\end{aligned}
%\label{eq:MW1}
\end{equation*}
and $\MS_{i}(j)$ is defined as follow:
\begin{itemize}
\item {\bf If} $j \bmod (2m_{i}-2) \neq (2m_{i}-3)$ and $j \bmod 2 = 0$, then $\MS_{i}(j) = {\GRlo}_{i}^{\big(j \bmod (2m_{i}-2)\big)/2}$; %${j/2 \bmod (m_{i}-1)}$
\item {\bf Else if} $j \bmod (2m_{i}-2) \neq (2m_{i}-3)$ and $j \bmod 2 = 1$, then $\MS_{i}(j) = {\CLlo}_{i}^{\big( (j \bmod (2m_{i}-2)) +1 \big)/2}$; %${(j+1)/2 \bmod m_{i}}$
\item {\bf Else if} $j = 2m_{i}-3$, then $\MS_{i}(j) = T_{i}-D_{i}+\CLlo_{i}^{m_i-1}+\CL_{i}^{0}$;
\item {\bf Else} $\MS_{i}(j) = T_{i}- \sum_{j=0}^{2m_i-3}\MLup_{i}^j -\sum_{j=1}^{m_i-2}\CLlo_{i}^j -\sum_{j=0}^{m_i-2}\GRlo_{i}^j$;
\end{itemize}
\end{lemma}
%\begin{equation*}
%\MS_{i}(j) = \left\{
%             \begin{array}{@{}l@{}}%{@{}l@{\mkern-36mu}r@{}}
%\displaystyle {\GRlo}_{i}^{j/2 \bmod m_{i}} \\ {\mkern+40mu}   \mbox{if\ } j \bmod 2m_{i}-2 \neq (2m_{i}-3) \mbox{\ and\ } j \bmod 2 = 0 \\
%\displaystyle {\CLlo}_{i}^{(j+1)/2 \bmod m_{i}} \\ {\mkern+40mu}   \mbox{if\ } j \bmod 2m_{i}-2 \neq (2m_{i}-3) \mbox{\ and\ } j \bmod 2 = 1 \\
%\displaystyle T_{i}-D_{i}+\CLlo_{i}^{m_i-1}+\CL_{i}^{0} {\mkern+50mu} \mbox{else if\ }  j = 2m_{i}-3  \\ %j \leq m_{i}
%\displaystyle T_{i}- \sum_{j=0}^{2m_i-3}\MLup_{i}^j -\sum_{j=1}^{m_i-2}\CLlo_{i}^j -\sum_{j=0}^{m_i-2}\GRlo_{i}^j \\ {\mkern+280mu}  \mbox{otherwise} 
%             \end{array}
%\right.
%\end{equation*}
\begin{proof}
From the perspective of executing memory copies over the bus, the $2m_{i}-2$ memory copy segments are the execution segments by the definition of self-suspension task in Section~\ref{sec:suspension}. So the definition of $\MW_{i}^{h}(t)$ and $l$ directly follows those in Lemma~\ref{lem:W} by applying $\MLup$ to $\Lup$ and changing from $m_i$ to $2m_{i}-2$.

The key difference is in the definition of $\MS_{i}(j)$, which is the minimum \enquote{interval-arrival time} between execution segments $\ML_i^j$ and $\ML_i^{j+1}$. By the RT-GPU task model, when $j \bmod (2m_{i}-2) \neq (2m_{i}-3)$, there is either a GPU or CPU segment after $\ML_i^j$, depending on whether the index is even or odd. So the lower bound on the response time of the corresponding GPU or CPU segment is the minimum {interval-arrival time} on the bus. For the latter case, the response time of a CPU segment is lower bounded by its minimum execution time. When $j = 2m_{i}-3$, $\ML_i^j$ is the last memory copy segment of the first job of $\tau_i$ occurring in the time interval $t$. In the worst case, all the segments of this job are delayed toward its deadline, so the minimum interval-arrival time between $\ML_i^j$ and $\ML_i^{j+1}$ is the sum of $T_{i}-D_{i}$, the minimum execution time of the last CPU segment $\CLlo_{i}^{m_i-1}$, and the minimum execution time of the first CPU segment $\CL_{i}^{0}$ of the next job. The last case calculates the minimum {interval-arrival time} between the last memory copy segment of a job that is not the first job and the first memory copy segment of the next job. Since these two jobs have an inter-arrival time $T_i$ between their first CPU segments, intuitively, $\MS_{i}(j)$ is $T_i$ minus all the segments of the previous job plus the last CPU segment $\CLlo_{i}^{m_i-1}$ of the previous job plus the first CPU segment $\CL_{i}^{0}$ of the next job, which is the above formula.
\end{proof}
\vspace{-2mm}
Hence, the response time of memory copy segment $\ML_k^j$ can be bounded by calculating the interference caused by the workload of tasks $hp(k)$ with higher-priorities than task $\tau_k$ and the blocking term from a low-priority task in $lp(k)$.  

\begin{lemma}
\label{lem:MRkj}
The worst-case response time $\MRup_{k}^{j}$ is the smallest value that satisfies the following recurrence:
\begin{equation}
\begin{split}
\MRup_{k}^{j} = \MLup_{k}^{j} &+ \sum_{\tau_{i}\in hp(k)}  \max_{h\in[0,2m_{i}-3]}{\MW_{i}^{h}}  (\MRup_{k}^{j}) \\
&+ \max_{\tau_{i}\in lp(k)} \max_{h\in[0,2m_{i}-3]} \MLup_{i}^{h}
\end{split}
\label{eq:MRkj}
\end{equation}
\end{lemma}
\begin{proof}
Because the execution of memory copy segments is non-preemptive, the calculation of $\MRup_{k}^{j}$ extends Lemma~\ref{lem:Rkj} by incorporating the blocking due to a low-priority memory copy segment that is already under execution on the bus. Under non-preemptive fixed-priority scheduling, a segment can only be blocked by at most one lower-priority segment, so this blocking term is upper bounded by the longest lower-priority segment.
\end{proof}
\vspace{-6mm}

\subsection{Fixed-Priority Scheduling for CPU Segments}
\label{sec:CR}

Now, we will switch the view and focus on analyzing the fixed-priority scheduling of the CPU segments. Looking from the perspective of the uniprocessor, CPU segments become the \enquote{execution segments}; the time intervals where task $\tau_i$ spends on waiting for memory copy and GPU to complete now become the \enquote{suspension segments}, since the processor can be used by other tasks during these intervals.

For now, let's assume that the upper bounds ${\MRup_{i}^{j}}$ and lower bounds $\MRlo_{i}^{j}$ on response times of memory copy segments are already given in Section~\ref{sec:MR}. As for GPU segments, the upper bounds $\GRup_{i}^{j}$ and lower bounds $\GRlo_{i}^{j}$ have been obtained in Section~\ref{sec:GR}. 
Similarly, we define the following CPU workload function $\CW_{i}^{h}(t)$. 

\begin{lemma}
\label{lem:CW}
$\CW_{i}^{h}(t)$ bounds the maximum amount of CPU computation that task $\tau_i$ can perform during an interval with a duration $t$ and a starting CPU segment $\CL_{i}^{h}$, where:
\begin{equation*}
\begin{aligned}
\CW_{i}^{h}(t) =& \sum_{j=h}^{l} \CLup_{i}^{j \bmod m_{i} } + \min \Big(\CLup_{i}^{(l+1) \bmod m_{i} },\\
&\qquad\qquad  t-\sum_{j=h}^{l} \big(\CLup_{i}^{j \bmod m_{i} }+\CS_{i}(j) \big) \Big)
\end{aligned}
%\label{eq:MW}
\end{equation*}
where $l$ is the maximum integer satisfying the following condition:
\begin{equation*}
\begin{aligned}
%\sum_{j=h}^{l}(ML_{i}^{j \bmod  m_{i}}+S_{i}^{j}) \leq t
\sum_{j=h}^{l} \big(\CLup_{i}^{j \bmod m_{i} }+\CS_{i}(j) \big) \leq t
\end{aligned}
%\label{eq:MW1}
\end{equation*}
and $\CS_{i}(j)$ is defined as follow:
\begin{itemize}
\item {\bf If} $j \bmod m_{i} \neq (m_{i}-1)$, then $\CS_{i}(j) = {\MLlo}_{i}^{2(j \bmod m_{i})}+{\GRlo}_{i}^{j \bmod m_{i}}+{\MLlo}_{i}^{2(j \bmod m_{i})+1}$;
\item {\bf Else if} $j = m_{i}-1$, then $\CS_{i}(j) = T_{i}-D_{i}$;
\item {\bf Else} $\CS_{i}(j) = T_{i}- \sum_{j=0}^{m_i-1}\CLup_{i}^j -\sum_{j=0}^{2m_i-3}\MLlo_{i}^j -\sum_{j=0}^{m_i-2}\GRlo_{i}^j$;
\end{itemize}
\end{lemma}
%\begin{equation*} %\displaystyle 
%\CS_{i}(j) = \left\{
%             \begin{array}{@{}l@{}}%{@{}l@{\mkern-36mu}r@{}}
%{\MLlo}_{i}^{2(j \bmod m_{i})}+{\GRlo}_{i}^{j \bmod m_{i}}+{\MLlo}_{i}^{2(j \bmod m_{i})+1} \\ {\mkern+50mu}   \mbox{if\ } j \bmod m_{i} \neq (m_{i}-1) \\
%T_{i}-D_{i} {\mkern+50mu} \mbox{else if\ }  j = m_{i}-1  \\ %j \leq m_{i}
%\displaystyle 
%T_{i}- \sum_{j=0}^{m_i-1}\CLup_{i}^j -\sum_{j=0}^{2m_i-3}\MLlo_{i}^j -\sum_{j=0}^{m_i-2}\GRlo_{i}^j \\ {\mkern+280mu}  \mbox{otherwise} 
%             \end{array}
%\right.
%\end{equation*}
\begin{proof}
From the perspective of the uniprocessor, the $m_{i}$ CPU segments are the execution segments by the definition of self-suspension task. So the definition of $\CW_{i}^{h}(t)$ and $l$ directly follows those in Lemma~\ref{lem:W} by applying $\CLup$ to $\Lup$.
For the minimum \enquote{interval-arrival time} $\CS_{i}(j)$, there are two memory copy and one GPU segments between segments $\CL_i^j$ and $\CL_i^{j+1}$ by the RT-GPU task model, when $j \bmod m_{i} \neq (m_{i}-1)$. So $\CS_{i}(j)$ is the sum of the minimum response times of these segments, where the response time of a memory copy segment is lower bounded by its minimum length. The case of $j = m_{i}-1$ is the same. The last case considers for a job that is not the first job in interval $t$. The calculation is similar to the one in Lemma~\ref{lem:W}, except that both the $2m_i-2$ memory copy and $m_i-1$ GPU segments constitute the suspension time.
%calculates the minimum {interval-arrival time} between its last CPU segment and the first CPU segment of the next job
\end{proof}
\vspace{-2mm}
Hence, the response time of CPU segment $\CL_k^j$ can be bounded by calculating the interference caused by the CPU workload of tasks $hp(k)$ with higher-priorities than task $\tau_k$.  
\vspace{-2mm}

\begin{lemma}
\label{lem:CRkj}
The worst-case response time $\CRup_{k}^{j}$ is the smallest value that satisfies the following recurrence:
\begin{equation}
\begin{split}
\CRup_{k}^{j} = \CLup_{k}^{j} &+ \sum_{\tau_{i}\in hp(k)}  \max_{h\in[0,m_{i}-1]}{\CW_{i}^{h}}  (\CRup_{k}^{j}) 
\end{split}
\label{eq:CRkj}
\end{equation}
\end{lemma}
\begin{proof}
The formula is directly extended from Lemma~\ref{lem:Rkj}.
\end{proof}

\vspace{-7mm}
\subsection{RT-GPU Scheduling Algorithm and Analysis}
\label{sec:algo}

For a particular virtual SM allocation $\VGN_i$ for all tasks $\tau_i$, we can calculate the response times of all GPU, memory copy, and CPU segments using formulas in Section~\ref{sec:GR} to \ref{sec:CR}. Note that a task starts with the CPU segment $\CL_{i}^{0}$ and ends with the CPU segment $\CL_{i}^{m_{i}-1}$. Therefore, we can upper bound the end-to-end response times for all tasks using the following theorem, by looking at the perspective from CPU.

\begin{theorem}
\label{lem:endRk}
The worst-case end-to-end response time $\Rup_{k}$ of task $\tau_k$ is upper bounded by the minimum of $\RupA_{k}$ and $\RupB_{k}$, i.e., $\Rup_{k} = \min (\RupA_{k}, \RupB_{k})$, where:
\begin{equation}
\RupA_{k} = \sum_{j=0}^{m_k-2}\GRup_{k}^j +\sum_{j=0}^{2m_k-3}\MRup_{k}^j + \sum_{j=0}^{m_{k}-1} \CRup_{k}^{j}
\label{eq:endRk1}
\end{equation}
and $R2_{k}$ is the smallest value that satisfies the recurrence:
\begin{equation}
\begin{split}
\RupB_{k} =& \sum_{j=0}^{m_k-2}\GRup_{k}^j +\sum_{j=0}^{2m_k-3}\MRup_{k}^j + \sum_{j=0}^{m_k-1}\CLup_{k}^j \\
&+ \sum_{\tau_{i}\in hp(k)} \max_{h\in[0,m_{i}-1]}{\CW_{i}^{h}} (\RupB_{k})
\label{eq:endRk2}
\end{split}
\end{equation}
\end{theorem}
\begin{proof}
The calculations for $\RupA_{k}$ and $\RupB_{k}$ are extended from Lemma~\ref{lem:Rk} by noticing that the time spent on waiting for GPU and memory copy segments to complete are {suspension segments} from the perspective of CPU execution.
\end{proof}
\vspace{-2mm}
With the upper bound on the response time of a task, the following corollary follows immediately.
\vspace{-2mm}
\begin{corollary}
\label{cor:endRk}
A CPU-GPU task $\tau_k$ is schedulable under federated scheduling on virtual SMs and fixed-priority scheduling on CPU and bus, if its worst-case end-to-end response time $\Rup_{k}$ is no more than its deadline $D_k$.
\end{corollary}
\vspace{-2mm}
\noindent\textbf{Computational complexity.} Note that the calculations for the worst-case response times of individual CPU and memory copy segments, as well as one upper bound on the end-to-end response time, involves fixed-point calculation. Thus, the above schedulability analysis has pseudopolynomial time complexity. \textcolor{black}{Given the system model notation in Section \ref{sec:task_model}, the grid search on spatial partitioning of $GN$ SMs has a complexity of $min(O({GN}^{n}),O(n^{GN}))$. The analysis of fixed-priority tasks on the memory copy and on the CPU have a complexity of $O(m_{i}^2)$ respectively. Therefore, the time complexity of the entire scheduling strategy with response time analysis is \vspace{-1mm}
\begin{equation}
\begin{aligned}
min(O({GN}^{n}m_{i}^4),O(n^{GN}m_{i}^4)).
\end{aligned}
\label{eq:complexity}
\end{equation}
}

\SetInd{0.5em}{0.5em}
\setlength{\algomargin}{0.5\parindent} 
\SetNlSkip{0.2em}
\begin{algorithm}
\caption{Fixed Priority Self-Suspension with Grid Searched Federated Scheduling} 
\label{Algorithm:schedulability_test}
{
\KwIn{Task set $\tau$, number of physical SMs $GN$}
\KwOut{Scheduability}
%\nl Scheduability $=0$\;
\ArgSty{//Generating enough virtual SMs:\\}
\For{$Thread \_ block \_ limit$ = 1024, 512, ...}{
\nl $\VGN = \frac{SM \_ thread \_ limit}{Thread \_ block \_ limit}$\\
\If{$\VGN \geq n$}{\textbf{Break};}
\ArgSty{//Grid search for federated scheduling of GPU segments:\\}
\nl \For{$\VGN_{1}$ = 1, ..., $\VGN$}{
    \nl \For{$\VGN_{i}$ = 1, ..., $\VGN-\sum_{j=1}^{i-1}\VGN_j$}{
        \nl \For{$\VGN_{n}$ = 1, ..., $\VGN-\sum_{j=1}^{n-1}\VGN_j$}{
\ArgSty{//Calculate response times of GPU segments:\\}  %lower and upper bounds  $\GRlo_{i}^{j}$ $\GRup_{i}^{j}$ 
\nl $\GRlo_i^j = \frac{\GWlo_i^{j}}{2\VGN_i},  1 \leq i \leq n$\;
\nl $\GRup_i^j = \frac{\GWup_i^{j}\alpha_i^j - \GLup_i^{j}}{2\VGN_i}+\GLup_i^{j},  1 \leq i \leq n$\;
\nl Calculate worst-case response time $\MRup_{k}^{j}$ for all memory copy segments using Eq.(\ref{eq:MRkj})\;
\nl Calculate worst-case response time $\CRup_{k}^{j}$ for all CPU segments using Eq.(\ref{eq:CRkj})\;
\nl Calculate worst-case end-to-end response time $\Rup_{k}$ for all tasks using Theorem~\ref{lem:endRk}\;
\nl \If{$\Rup_{k} \leq D_k$ for all $\tau_k$}{\nl Scheduability $=1$; break out of all for loops\;}
        }
    }
}
}
}
\end{algorithm}
\vspace{-5mm}

Note that the above schedulability analysis assumes a given virtual SM allocation under federated scheduling. Hence, we also need to decide the best virtual SM allocation for task sets, in order to get better schedulability. The following RT-GPU Scheduling Algorithm adopts a brute force approach to deciding virtual SM allocation. Specifically, it enumerates all possible allocations for a given task set on a CPU-GPU platform and uses the schedulability analysis to check whether the task set is schedulable or not. Alternatively, a greedy approach can be applied, if one needs to reduce the running time of the algorithm while a slight loss in schedulability is affordable. Given the number of SMs assigned to the tasks and the CPU and GPU execution time, the schedulability under current resource utilization rate can be calculated following the procedure from subsection \ref{sec:GR} to subsection  \ref{sec:algo}. 
The full procedure of scheduling GPU tasks can be described as follows: (1) Grid search \cite{chen2014fixed} a federated scheduling for the GPU codes and calculate the GPU segment response time $[\GRlo_{i}^{j} \; \GRup_{i}^{j}]$, details in Section \ref{sec:schedule}.4. (2) The CPU segments and memory copy segments are scheduled by fixed priority scheduling. (3) If all the tasks can meet the deadline, then they are schedulable and otherwise go back to step (1) to grid search for the next federated scheduling. This schedulability test can be summarized with pseudo code in Algorithm 2.
\vspace{-6mm}

\textcolor{black}{
\subsection{Roadmap of Extending the Scheduling}
\vspace{-1mm}
Moreover, this end-to-end response time analysis is not limited to CPU-memory-GPU systems. It can also be directly applied to other heterogeneous systems with one type of heterogeneous core, like CPU-memory-FPGA and CPU-memory-TPU systems. To schedule the systems with multiple GPUs (with the same type of GPU SMs), a new constraint must be added to the GPU SM allocation part (after line 3 in Algorithm 2). In this constraint, the GPU SM allocation can only be valid when all the tasks are not executed on the SMs that belong to different GPUs. Further to schedule the systems with multiple GPUs (with different types of GPU SMs), the above constraint must be added. Also, the lower and upper bounds of GPU segment response time (line 4 and line 5 in Algorithm 2) must be calculated with the corresponding computing power of the different types of GPU SMs. After this updated GPU federated scheduling, the fixed-priority scheduling of CPU and memory copy can be directly applied.
}
\vspace{-4mm}

\section{Full-System Evaluation}

We now present an evaluation of our approach.  Section~\ref{sec:experiments} describes experiments we conducted to validate our approach.
Section~\ref{sec:implementation} explains how we implemented persistent threading and workload pinning in those experiments.  Section~\ref{sec:analysis} discusses our analytical evaluations of schedulability under our approach.
Finally, Section~\ref{sec:real} presents schedulability results on real GPU systems.
\vspace{-2mm}

\subsection{Experiments}
\label{sec:experiments}
We conducted extensive experiments to evaluate the performance of the proposed RTGPU real-time scheduling approach. 
We choose self-suspension~\cite{bletsas2015errata}, STGM~\cite{saha2019stgm}: Spatio-Temporal GPU Management for Real-Time Tasks, and Enhanced MPCP \cite{patel2018analytical}: Analytical enhancements and practical insights for mpcp with self-suspensions as baselines to compare with, as they represent the state-of-the-art in both entire GPU and fine-grained (SM-granularity) GPU real-time scheduling algorithms and schedulability tests. 
\textbf{1. Proposed RTGPU:} the proposed real-time GPU scheduling of hard deadline parallel tasks with fine-grain utilization of persistent threads, interleaved execution, virtual SM, and fixed-priority federated scheduling.
\textbf{2. Self-Suspension:} real-time GPU scheduling of hard deadline parallel tasks with the persistent threads with self-suspension scheduling, as in \cite{bletsas2015errata}.
\textbf{3. STGM:} real-time GPU scheduling of hard deadline parallel tasks with the persistent threads and busy-waiting scheduling, as in \cite{saha2019stgm}. This work also tested and analyzed the self-suspension scheduling under different scenarios.
\textbf{4. Enhanced MPCP:} real-time GPU scheduling of hard deadline parallel tasks with hybrid approach of the enhancements and practical insights for MPCP with self-suspension, as in \cite{patel2018analytical}.
Please note that there is no previous scheduling algorithm that exactly matches the proposed system model. All the above scheduling algorithms are modified to match the proposed model. Some unique good features may be slightly scarified in the modification. For example, STGM can support multiple CPU cores but in our system model, we only use one CPU core.

\begin{figure*}
\setlength{\abovecaptionskip}{-0.02cm}
\centering
\subfigure[computation:suspension=2:1]{
\label{fig:fig_schedulability_length_1} %% label for second subfigure
\includegraphics[width=0.3\textwidth]{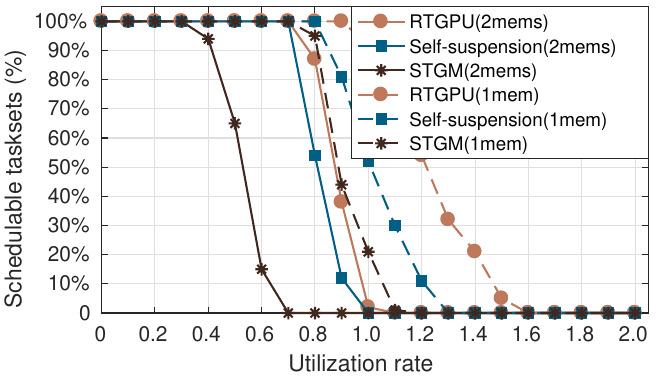}}
\subfigure[computation:suspension=1:2]{
\label{fig:fig_schedulability_length_2} %% label for second subfigure
\includegraphics[width=0.3\textwidth]{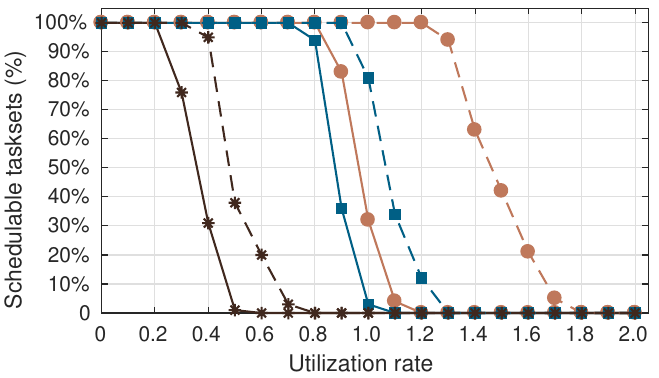}}
\subfigure[computation:suspension=1:8]{
\label{fig:fig_schedulability_length_3} %% label for second subfigure
\includegraphics[width=0.3\textwidth]{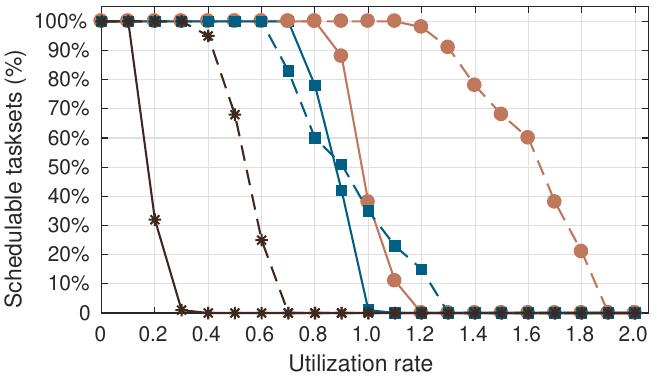}}
\caption{Schedulability under different computation (CPU) and suspension (memory+GPU) lengths}
\label{fig:schedulability_length} %% label for entire figure
\vspace{-0.5cm}
\end{figure*}

\begin{figure*}
\setlength{\abovecaptionskip}{-0.02cm}
\centering
\subfigure[3 subtasks]{
\label{fig:schedulability_subtask_1} %% label for second subfigure
\includegraphics[width=0.3\textwidth]{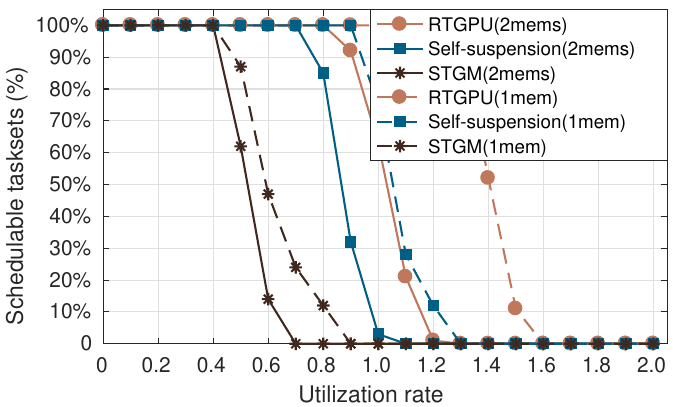}}
\subfigure[5 subtasks]{
\label{fig:schedulability_subtask_2} %% label for second subfigure
\includegraphics[width=0.3\textwidth]{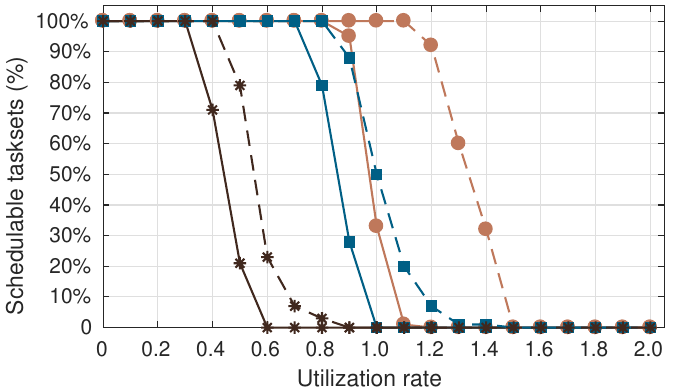}}
\subfigure[7 subtasks]{
\label{fig:schedulability_subtask_3} %% label for second subfigure
\includegraphics[width=0.3\textwidth]{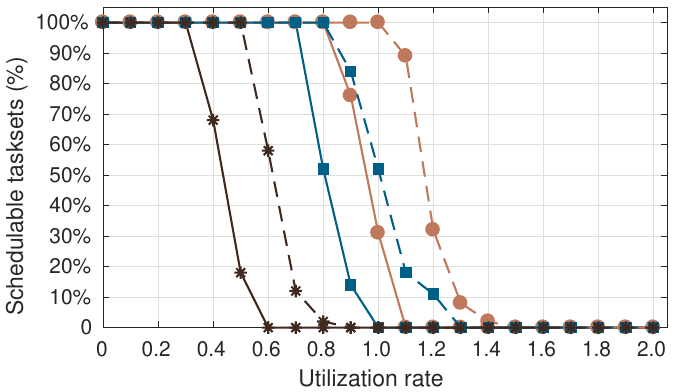}}
\caption{Schedulability under different numbers of subtasks}
\label{fig:schedulability_subtask} %% label for entire figure
\vspace{-0.5cm}
\end{figure*}

To compare the schedulability results for these approaches, we measured the acceptance ratio in each of four simulations with respect to a given goal for taskset utilization. We generated 100 tasksets for each utilization level, with the following task configurations. The acceptance ratio of a level was the number of schedulable tasksets, divided by the number of tasksets for this level, i.e., 100. According to the GPU workload profiling and characterization \cite{chen2017gaas}, the memory length upper bound was set to 1/4 of the GPU length upper bound. We first generated a set of utilization rates, $U_i$, with a uniform distribution for the tasks in the taskset, and then normalized the tasks to the taskset utilization values for the given goal. Next. we generated the CPU, memory, and GPU segment lengths, uniformly distributed within their ranges in Table \ref{tab:taskset_generation}. The deadline $D_{i}$ of task $i$ was set according to the generated segment lengths and its utilization rate: $D_{i} = (\sum_{j=0}^{m_{i}-1} \CLup_{i}^{j} + \sum_{j=0}^{2m_{i}-3} \MLup_{i}^{j} + \sum_{j=0}^{m_{i}-2} \GLup_{i}^{j})/U_{i}$. 
In the configuration setting, the CPU, memory, and GPU lengths were normalized with one CPU, one memory interface, and one GPU SM. When the total utilization rate, $U$, is 1, the one CPU, one memory interface, and one GPU SM are fully utilized. As there are multiple SMs available (and used), the total utilization rate will be larger than 1. The period $T_{i}$ is equal to the deadline $D_{i}$. The task priorities are determined with deadline-monotonic priority assignment.
Meanwhile, in each experiment we evaluate two models. The first model has two memory copies: one memory copy from CPU to GPU and one memory copy back from GPU to CPU between a CPU segment and a GPU segment, which is exactly the execution model we introduced in section \ref{sec:GPU_model}. The second model has one memory copy between a CPU segment and a GPU segment, which combines the memory copy from CPU to GPU and the memory copy from GPU to CPU. These two models can capture not only the CPU-GPU systems but also general heterogeneous computing architectures.
\vspace{-2mm}

\begin{table}[t]
\caption{Parameters for the taskset generation}
\setlength{\belowcaptionskip}{-4mm}
\label{tab:taskset_generation}
\begin{tabular}{|l|c|c|}
\hline  
\textbf{Parameters} & \textbf{Value}\\ 
\hline
Number of tasks $N$ in taskset & 5 \\ 
\hline
Task type & periodic tasks \\
\hline
Number of subtasks $M$ in each task & 5 \\ 
\hline
Number of tasksets in each experiment & 100 \\ 
\hline
CPU segment length (ms) & [1 to 20] \\ 
\hline
Memory segment length (ms) & [1 to 5] \\ 
\hline
GPU segment length (ms)  & [1 to 20] \\ 
\hline
Task period and deadline & $(T_{i}/D_{i})$ \\ 
\hline
Number of physical GPU SMs $N_{SM}$ & 10 \\ 
\hline
Priority assignment & D monotonic \\ 
\hline
\end{tabular}
\vspace{-2mm}
\end{table}

%\vspace{-5mm}
\subsection{System Side Implementation}
\label{sec:implementation}
First on the system side, we implement the persistent thread with workload pinning and self-interleaving. As studied in Section \ref{sec:GPU_model}, the workload pinning and self-interleaving can improve the system throughput under GPU fine-grain partitioning. We further test the workload pinning and self-interleaving with five real benchmarks from the Rodinia and CUDA SDK tasksets representing different but common types of GPU applications. Table \ref{tab:interleaved_ratio} shows the interleaved execution ratios measured from these application with different SMs assigned to each kernel. As the kernel is interleaved with itself, each kernel has a relatively stable interleaved execution ratio when different numbers of SM are assigned to the kernel. Since the benchmark is divided into two pieces and these two pieces are executed simultaneously on the given number of SMs, the throughput improvements can be calculated with $2/{\alpha}$. Therefore, the benchmarks achieve 11\% to 81\% throughput improvements on benchmark Vectoradd and Quasirand. Quasirand achieves significant throughput improvement because the original Quasirand uses less than half of the SM resources. Self-interleaving can leverage the remaining resources to achieve throughput improvement. Next, we will evaluate the schedulability of the proposed approach. To have a fair comparison, the following experiments are all based on the workload pinning and self-interleaving.
\vspace{-4mm}

\begin{table}[t]
\caption{Interleaved execution ratios measured from real benchmarks with different SMs assigned to each kernel}
\setlength{\belowcaptionskip}{-4mm}
\label{tab:interleaved_ratio}
\begin{tabular}{|c|c|c|c|c|}
\hline  
\textbf{Benchmark/Number of SMs} & \textbf{1} & \textbf{2} & \textbf{4} & \textbf{8}\\ 
\hline
\textbf{Dxtc} & 1.68 & 1.66 & 1.69 & 1.64\\ 
\hline
\textbf{BFS} & 1.56 & 1.61 & 1.59 & 1.57\\ 
\hline
\textbf{Particle Filter} & 1.42 & 1.45 & 1.41 & 1.46\\ 
\hline
\textbf{Vectoradd} & 1.80 & 1.77 & 1.80 & 1.78\\ 
\hline
\textbf{Quasirand} & 1.22 & 1.23 & 1.24 & 1.23\\ 
\hline
\end{tabular}
\vspace{-2mm}
\end{table}

\subsection{Schedulability Analysis}
\label{sec:analysis}
Our analytical evaluation focused on the schedulability of tasksets as the overall utilization increased, with respect to different parameters pertinent to schedulability.  
The following sub-subsections present the results of four simulations that each varied the different parameters we examined: the ratios of CPU, memory, and GPU segment lengths; the number of subtasks; the number of tasks; and the number of total SMs. 
\vspace{-3mm}

\subsubsection{CPU, Memory, and GPU Lengths}
We investigated the impact of CPU, memory, and GPU segment lengths on the acceptance ratio. To study this quantitatively, We tested the acceptance ratio under different length range ratios. The CPU length is shown as Table \ref{tab:taskset_generation} and we changed the memory, and GPU lengths according to the length ratio. 
Fig. \ref{fig:schedulability_length} shows taskset acceptance ratio when the CPU, memory, and GPU length range ratios were set to 2:1, 1:2, and 1:8, which give an exponential scale.

\begin{figure*}
\setlength{\abovecaptionskip}{-0.02cm}
\centering
\subfigure[3 tasks]{
\label{fig:schedulability_task_1} %% label for second subfigure
\includegraphics[width=0.3\textwidth]{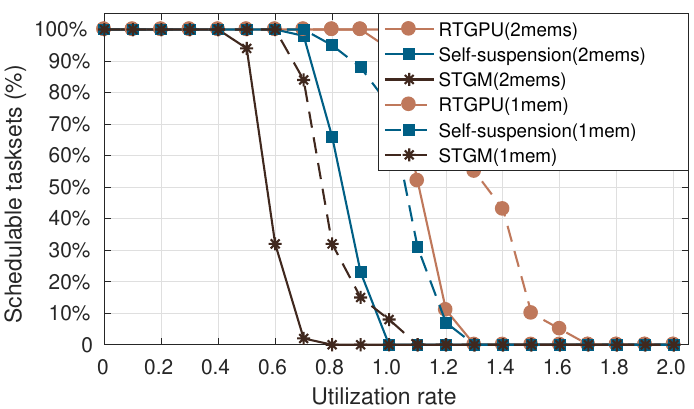}}
\subfigure[5 tasks]{
\label{fig:schedulability_task_2} %% label for second subfigure
\includegraphics[width=0.3\textwidth]{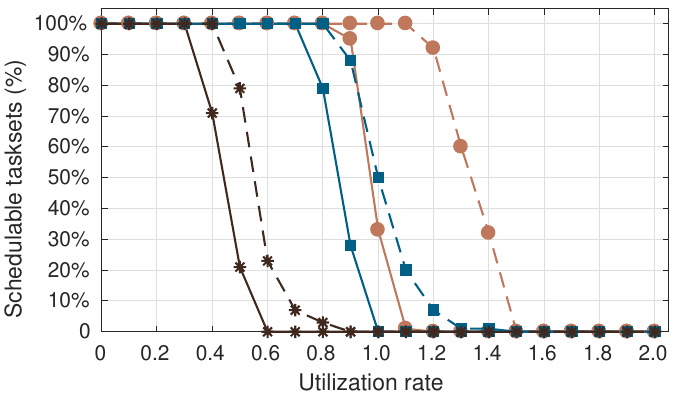}}
\subfigure[7 tasks]{
\label{fig:schedulability_task_3} %% label for second subfigure
\includegraphics[width=0.3\textwidth]{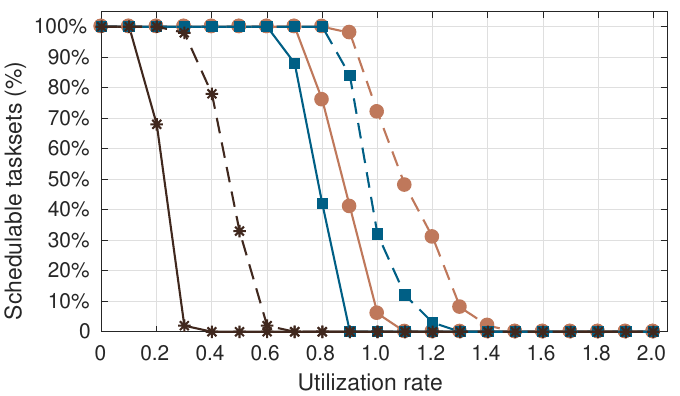}}
\caption{Schedulability under different numbers of tasks}
\label{fig:schedulability_task} %% label for entire figure
\vspace{-0.5cm}
\end{figure*}
\begin{figure*}
\setlength{\abovecaptionskip}{-0.02cm}
\centering
\subfigure[5 SMs]{
\label{fig:schedulability_sm_1} %% label for second subfigure
\includegraphics[width=0.3\textwidth]{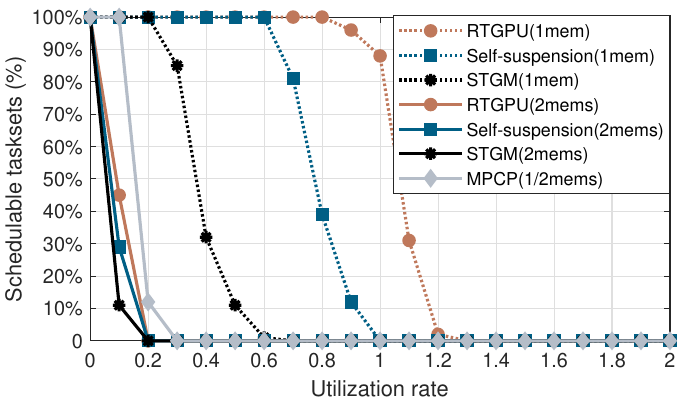}}
\subfigure[8 SMs]{
\label{fig:schedulability_sm_2} %% label for second subfigure
\includegraphics[width=0.3\textwidth]{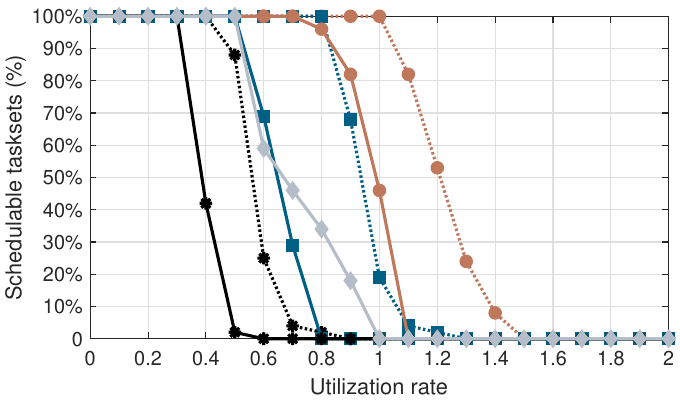}}
\subfigure[10 SMs]{
\label{fig:schedulability_sm_3} %% label for second subfigure
\includegraphics[width=0.3\textwidth]{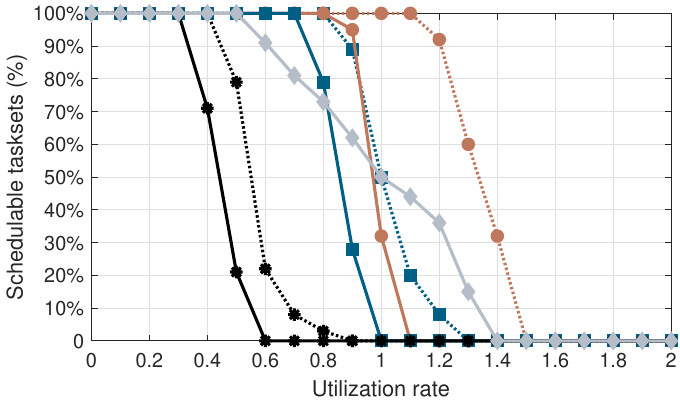}}
\caption{Schedulability under different numbers of SMs}
\label{fig:schedulability_sm} %% label for entire figure
\vspace{-0.5cm}
\end{figure*}

Not surprisingly, the STGM approach is effective only when the memory and GPU segment (suspension segment) lengths are sufficiently short: the STGM approach was developed based on ”busy waiting”. When tasks are being processed in memory copy and GPU segments, the CPU core is not released and remains busy waiting for the memory copy and GPU segments to finish. Although this is the most straightforward approach, its pessimistic aspect lies in the CPU waiting for the memory copy and GPU segments to finish. Thus, it will be ineffective and hugely pessimistic when the memory copy and GPU segments are large.

Self-suspension scheduling in \cite{schonberger2018schedulability} increases the schedulability performance compared with the straight forward STGM approach. Self-suspension models the memory and GPU segments as being suspended, and the CPU is released during this suspension. The theoretical drawback of this approach is that the suspension does not distinguish between the memory segments and GPU segments. Instead, they are modelled as non-preemptive and will block higher priority tasks. However, in real systems, each task is allocated its own exclusive GPU SMs, and the GPU segments in one task will not interfere the GPU segments in other tasks.

Enhanced MPCP in \cite{patel2018analytical} has the best performance when the CPU segments are large. The schedulability is sensitive to the CPU, memory, and GPU segment lengths ratios. It will decrease obviously when the GPU segments become longer because Enhanced MPCP is designed for the heterogeneous systems with multiple CPU cores which targets the applications with long CPU segments.

The RTGPU schedulability analysis proposed in this paper is effective even when the memory and GPU segment (suspension segment) lengths are long. For example, in the 1:8 length test, RTGPU reaches 1.1 utilization rate with 100\% schedulability, which is 57\% improvement with previous work. In this approach, we distinguish the CPU, memory, and GPU segments based on their individual properties. For example, if the CPU cores are preemptive, then no blocking will happen. Blocking happens only in non-preemptive memory segments. Meanwhile, because federated scheduling is applied for the GPU segments and each task is allocated its own exclusive GPU SMs, the GPU segments can be executed immediately when they are ready, without waiting for higher priority GPU segments to finish or being blocked by lower GPU segments.

Also, by comparing the models with one memory copy and two memory copies, we notice that the memory copy is the bottleneck in the CPU-GPU systems because of limited resource (bandwidth) and non preemption. Reducing the numbers of memory copies or combining memory copies can increase the system schedulability, especially when the memory copy length is large shown in Fig. \ref{fig:schedulability_length} (b) and (c).
\vspace{-3mm}

\subsubsection{Number of Subtasks}
We then evaluated the impact of the number of subtasks in each task on the acceptance ratio. From the possible values in Table \ref{tab:taskset_generation}, the number of subtasks, $M$, in each task was set to 3, 5, or 7. The corresponding acceptance ratios are shown in Fig.\ref{fig:schedulability_subtask}. The results show that with more subtasks in a task, schedulability decreases under all approaches but the proposed RTGPU approach still outperforms all other approaches. The Enhanced MPCP approach is the most robust scheduling algorithm as the  subtasks increase.
\vspace{-3mm}

\subsubsection{Number of Tasks}
In a third simulation, we evaluated the impact of the number of tasks in each taskset on the acceptance ratio. Again, from the possible values in Table \ref{tab:taskset_generation}, the number of tasks, $N$, in each task was set to 3, 5, or 7. The corresponding acceptance ratios are shown in Fig.\ref{fig:schedulability_task}. As with subtasks, schedulability decreases under all the approaches as the number of tasks increases, but the proposed RTGPU approach outperformed the other two.
\vspace{-3mm}

\subsubsection{Number of SMs}
Finally, we examined the impact of the number of total SMs on the acceptance ratio. Based on the possible values in Table \ref{tab:taskset_generation}, the number of subtasks M and tasks N in each setting are again set to 5. The corresponding acceptance ratios are shown in Fig.\ref{fig:schedulability_sm}. All approaches have better schedulability as the number of available SMs increases. From this set of experiments we can see that adding two more SMs will cause the utilization rate to increase for all approaches. Meanwhile, among all the approaches, the proposed RTGPU approach again achieves the best schedulability across different numbers of SMs. The schedulability of the Enhanced MPCP approach also increases significantly with the increased number of SMs. When the number of SMs is large enough, the Enhanced MPCP approach also performs well. As shown in Fig.\ref{fig:schedulability_sm} (a), when the computation resources (GPU SMs) are limited, the bottleneck from memory copy is more obvious and serious. The two memories model has a poor scheduability in all approaches and the one memory model has a significant improved performance.
\vspace{-4mm}

\subsection{Schedulability on Real GPU Systems}
\label{sec:real}
\textcolor{black}{To test and compare schedulability between the theoretical boundary and the performance on real GPU systems, we empirically evaluated the proposed RTGPU scheduling framework on an NVIDIA 1080TI GPU, under enough and a limited number of SMs.} The CPU was an Intel(R) Core(TM) i7-3930K CPU operating at 3.20GHz. We implemented the synthetic benchmarks described in Section \ref{sec:GPU_model} in a common real-time scheduling context, since multiple GPU kernel concurrency is supported only within the same CUDA context. 
To avoid interference from adaptive power setting and guarantee hard deadlines, we manually fixed the SM core and memory frequencies respectively using the nvidia-smi command. We also set the GPUs to persistence mode to keep the NVIDIA driver loaded even when no applications are accessing the cards.

\begin{figure}
\centering
\includegraphics[width=0.5\textwidth]{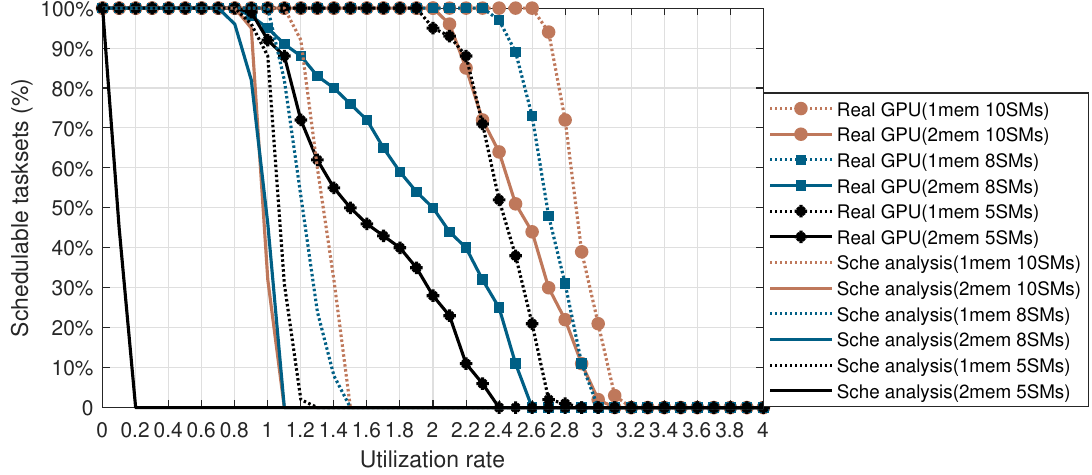}
\caption{Full system schedulability of 5 parallel tasks under 5, 8, 10 physical SMs}
\label{fig:schedulability_GPU_worst}
\vspace{-4mm}
\end{figure}

\begin{figure}
\centering
\includegraphics[width=0.5\textwidth]{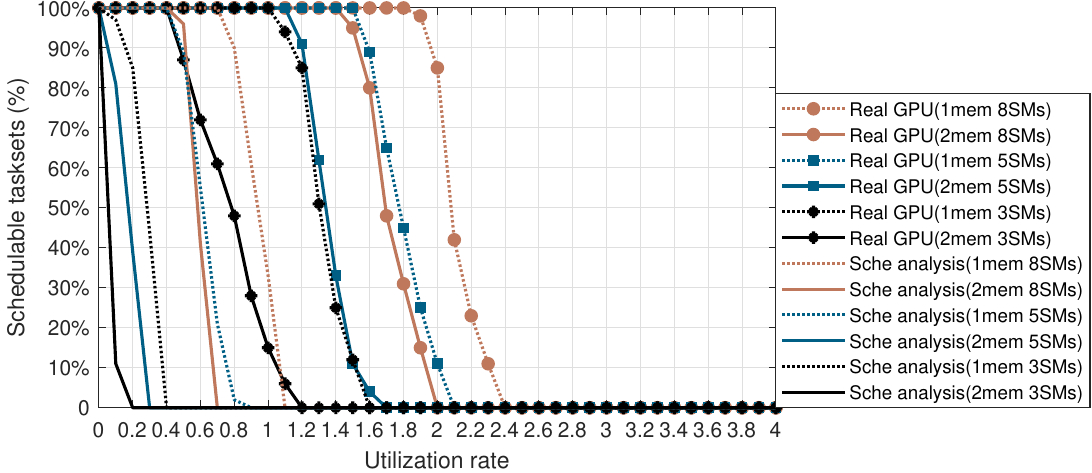}
\caption{Full system schedulability of 12 parallel tasks under 3, 5, 8 physical SMs}
\label{fig:schedulability_GPU_limited_SM}
\vspace{-5mm}
\end{figure}

As in the previous schedulability analysis experiments, each task in a taskset was randomly assigned one of the values in Table~\ref{tab:taskset_generation}. The deadline was set to the same value as the period.
\textcolor{black}{In this work, we use a measurement-based analysis to get the values of the kernel model (including the critical-path overhead). The execution time distributions of different sizes of memory copies through PCIe from CPU to GPU and from GPU to CPU and different GPU kernel thread lengths are measured from 10,000 samples.} Based on the measured worst case execution time, we calculate the values of the kernel model and build the worst execution time model. Then using the real GPU system, we examined schedulability using different numbers of SMs and compared the results from the schedulability analysis and from the real GPU experiments. Fig. \ref{fig:schedulability_GPU_worst} presents the acceptance ratio results of the RTGPU schedulability analysis and experiments on the real GPU system. Both of them have better schedulability as the number of available SMs increases. The gaps between the schedulability analysis and real GPU system arise from the pessimistic aspect of the schedulability analysis and the model mismatches between worst execution time and actual execution time. In the limited computation resource scenarios (5 SMs and 8 SMs), the bottlenecks from memory copy exist in both schedulability test and experiments with real GPU systems. Reducing the numbers of memory copies or combining memory copies are proper methods to deal with the bottlenecks. 

\textcolor{black}{Finally, we evaluate the system schedulability when it has a limited number of physical SMs (i.e., the number of physical SMs is smaller than the number of tasks). Fig. \ref{fig:schedulability_GPU_limited_SM} presents the system schedulability and corresponding response time analysis under 12 parallel tasks with 3, 5, and 8 physical SMs. To let each task have at least one allocated virtual SM, we generate four virtual SMs from every physical SM. Not surprisingly, the system suffers a lower schedulability when there are more parallel tasks and fewer GPU SMs. If we compare the schedulability from different numbers of parallel tasks or different numbers of physical SMs, the number of physical SMs has a more impact on the schedulability because the number of physical SMs directly determines the full system utilization rate.}

%the memory copy and GPU kernels are modeled by their average execution times. The results from the RTGPU schedulability analysis and real GPU system are presented in Fig.\ref{fig:schedulability_GPU_average}. Because the segments are modeled by their average execution times, which is much tighter than the worst execution time, the gaps between the schedulability analysis and experiments on the real GPU system are further reduced.
\vspace{-3mm}
\section{Related Work}
\vspace{-1mm}
For real-time systems with GPUs, previous work mainly leveraged GPU kernel-granularity scheduling. For example, Kato \cite{kato2011timegraph} introduced a priority-based scheduler. Elliott proposed shared resources and containers for integrating GPU and CPU scheduling \cite{elliott2012globally} and GPUSync \cite{elliott2013gpusync} for managing multi-GPU multicore soft real-time systems with flexibility, predictability, and parallelism. 
Golyanik \cite{golyanik2017towards} described a scheduling approach
based on time-division multiplexing; $S^3$DNN \cite{zhou2018s} optimized the execution of DNN GPU workloads in a real-time multi-tasking environment through scheduling the GPU kernels. Thermal and energy efficieny in GPU real-time scheduling systems were studied in \cite{santriaji2018merlot,hosseinimotlagh2019thermal,santriaji2018merlot}. However, these approaches focus on predictable GPU control, and 
hard to support multiple tasks to use the GPU at the same time. Thus, the GPU may be underutilized and a task may wait a long time to access the GPU.

\textcolor{black}{In the past few years, GPU vendors and researchers started to provide a more flexible GPU kernel execution manner such as temporal preemption and spatial partitioning.} \textcolor{black}{For example, NVIDIA started the initial preemption since Pascal architecture and in the recent Tegra architecture for embedded systems, the preemption types can be selected by users according to the application priorities. Besides, researchers also developed many frameworks to further support GPU preemption at kernel or even finer granularity.
For example, Park \cite{park2015chimera}, Basaran \cite{basaran2012supporting}, Tanasic \cite{tanasic2014enabling}, and Zhou \cite{zhou2015gpes} proposed architecture extensions with hardware and software codesigns to improve the preemption and tested on the GPU simulators. \textcolor{black}{Capodieci \cite{capodieci2018deadline} further presented a deadline-based real-time scheduling with the support of preemption on a NVIDIA Drive-PX GPU. The scheduler runs as a software partition on top of the NVIDIA hypervisor and leverages pixel-level preemption and thread-level preemption. This preemptive execution pattern implements and tests a preemptive Earliest Deadline First (EDF) scheduler. Extensive experiments demonstrate that preemptive EDF scheduling achieves significant schedulability improvement.
}
The Effisha framework~\cite{chen2017effisha} introduced software techniques without any hardware modification to support kernel preemption at the end of any arbitrary thread block.}
\textcolor{black}{
Meanwhile, targeting embedded systems without hardware nor driver stack extensions, Hartmann \cite{hartmann2019gpuart} also developed a fixed point preemption on GPUs called GPUart and evaluated the Gang-Earliest Deadline First and Gang-Fixed Task Priority scheduling strategies on it. According to the experimental results, up to 221x response time improvements are achieved in GPUart.}

On spatial partitioning, NVIDIA launched the MPS and MiG process management software to manage kernel parallel execution. AMD released open-source software support for hardware partitioning which has the potential to accelerate and aid the long-term viability of real-time GPU research \cite{otterness2021exploring,otterness2020amd}. Researchers \cite{gupta2012study,yu2018smguard,wu2015enabling} proposed the persistent thread techniques as discussed in Section \ref{background}. \textcolor{black}{Following the persistent thread technique, \cite{wang2021balancing} presents an energy-efficient scheduler sBEET by partitioning the computing resources and isolating kernel execution. Experiments on NVIDIA Jetson Xavier AGX demonstrate the sBEET could reduce deadline misses and energy consumption by up to 13\% and 21\%.} Liang \cite{liang2014efficient} introduced a software-hardware solution for efficient spatial-temporal multitasking for GPU. However, the computation throughput \cite{kayiran2014managing,yang2011hybrid} is usually the focus of GPU spatial partitioning. \cite{saha2019stgm,lee2014improving} considers the fine-grained real-time GPU scheduling only with the state-of-the-art system side work (persistent threads) and scheduling analysis. However, SM-granularity resource partitioning without an efficient real-time scheduling algorithm is not sufficient to achieve effective SM-Level scheduling with fine granularity and high utilization rates.
\textcolor{black}{According to the related works on temporal access and spatial portioning, temporal access based on preemption has a more flexible GPU access but spatial partitioning could achieve a higher schedulability with the additional requirement that the number of virtual SMs should be larger than the number of parallel tasks.}

Although flexible task execution can improve system schedulability, rare work provides a complete solution, which can seamlessly link the system improvement with efficient real-time scheduling algorithms. To obtain more universal and effective real-time GPU scheduling, and to piggyback on previous work, we propose real-time GPGPU scheduling: RTGPU. 
%Our method leverages architecture information to support finer-grain SM-level scheduling and to accurately model workload timing behavior with a virtual SM model. 
Compared with previous work, RTGPU leverages architecture information to support finer-grain SM-level scheduling and improves the schedulability and increases the throughput of real-time GPU systems.\vspace{-4mm}

\section{Conclusion}
\label{sec:conclusion}
To execute multiple parallel real-time applications on GPU systems, we propose \textit{RTGPU}---a real-time scheduling method including both system work and and a real-time scheduling algorithm with schedulability analysis. RTGPU leverages a precise timing model of the GPU applications with the persistent threads technique and achieves improved fine-grained utilization through interleaved execution.
The RTGPU real-time scheduling algorithm is able to provide real-time guarantees of meeting deadlines for GPU tasks with better schedulability compared with previous work.
We empirically evaluate our approach using synthetic benchmarks both via schedulability analysis and on real NVIDIA GTX1080Ti GPU systems, the results of which demonstrate significant performance gains compared to existing methods.
\vspace{-5mm}

%\IEEEpubidadjcol

%\ifCLASSOPTIONcompsoc
%\else
%\fi


\ifCLASSOPTIONcaptionsoff
  \newpage
\fi

\bibliographystyle{unsrt}
\bibliography{bibliography}

\begin{IEEEbiography}[{\includegraphics[width=1in,height=1.25in,clip,keepaspectratio]{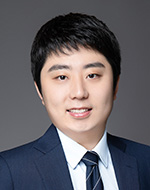}}]{An Zou}
is an Assistant Professor at the University of Michigan-Shanghai Jiao Tong University Joint Institute in the Shanghai Jiao Tong University. His research focuses on computer architecture and embedded systems. He broadly investigated a set of techniques and solutions via a bottom-up layered approach to improve computing power and performance efficiency. Dr. An Zou received his Ph.D. degree in Electrical Engineering from Washington University in St. Louis in 2021 and his M.S. and B.S. degrees from Harbin Institute of Technology (HIT) in 2015 and 2013. His work has been extensively published and recognized at top-tier conferences and journals including MICRO, DAC, ICCAD, AAAI, TCAD, TACO, and RTAS. He was a recipient of A. Richard Newton Young Student Fellow Award and the Best Paper Nominations at DAC 2017 and MLCAD 2020. 
\end{IEEEbiography}
\vspace{-15mm}

\begin{IEEEbiography}[{\includegraphics[width=1in,height=1.25in,clip,keepaspectratio]{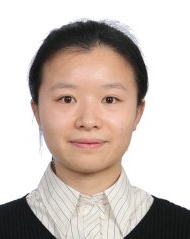}}]{Jing Li} 
is an Assistant Professor in Department of Computer Science at New Jersey Institute of Technology. She received her Ph.D. at Washington University in St. Louis in 2017, where she was advised by Professor Chenyang Lu and Kunal Agrawal. She received B.S. in computer science from Harbin Institute of Technology in 2011. Her research interests include real-time systems, parallel computing, cyber-physical systems, and reinforcement learning for system design and optimization. She has high impact publications in top journals and conferences with 3 outstanding papers. 
\end{IEEEbiography}
\vspace{-15mm}

\begin{IEEEbiography}[{\includegraphics[width=1in,height=1.25in,clip,keepaspectratio]{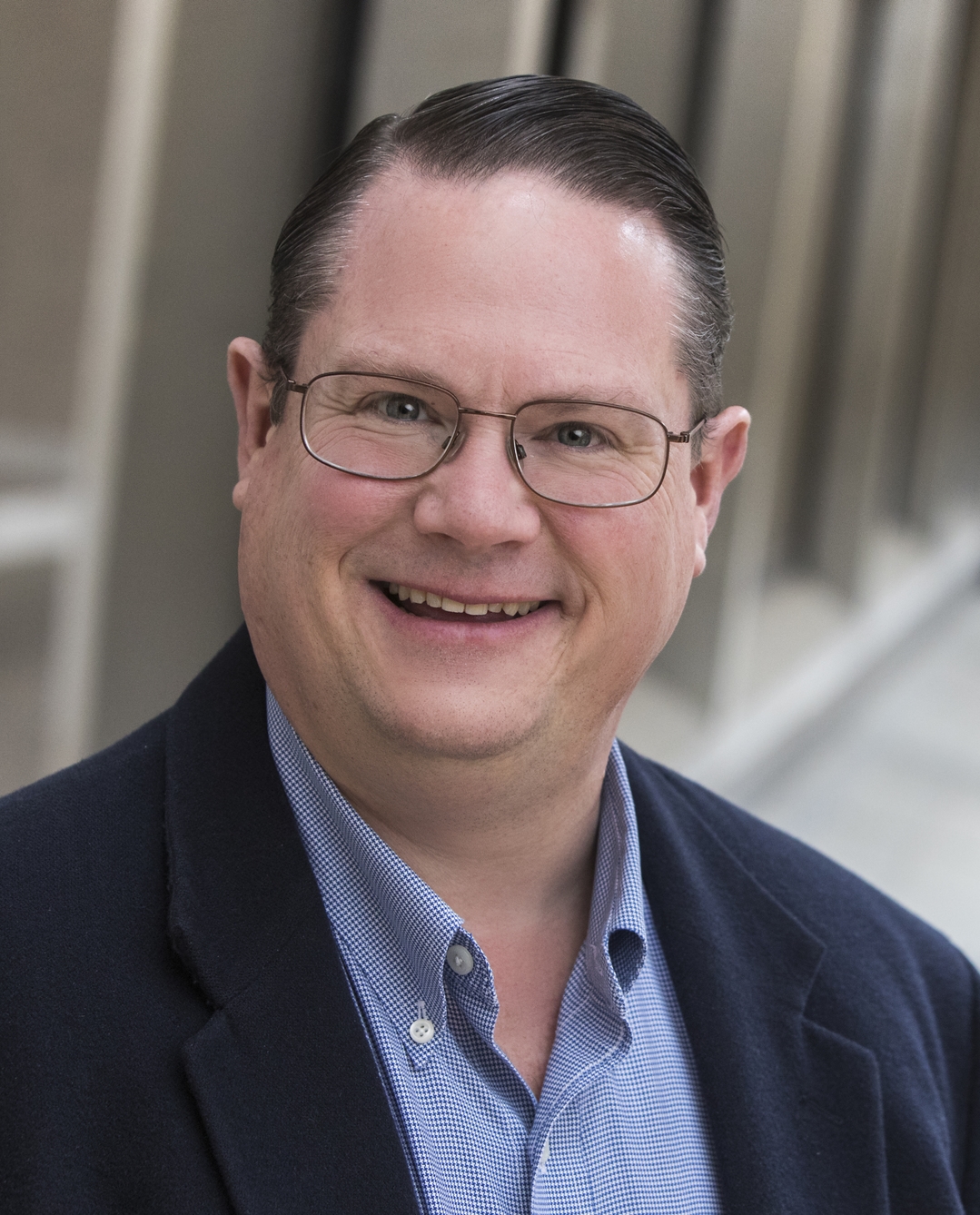}}]{Christopher D. Gill}
is a Professor in the Department of Computer Science and Engineering at Washington University in St. Louis. He has published more than 100 technical articles in selective peer-reviewed conferences and journals, and has led or contributed to the development, evaluation, and open-source release of numerous real-time systems research platforms and artifacts, including: the Kokyu real-time scheduling and dispatching framework that was used in several AFRL and DARPA projects and flight demonstrations; the nORB small-footprint real-time object request broker; a number of real-time and fault-tolerant services for The ACE ORB (TAO) and the Component Integrated ACE ORB (CIAO); the Cyber-physical Instrument for Real-time hybrid Structural Testing (CIRST) that established key foundations for real-time hybrid simulation (RTHS), and the CyberMech platform that built on the CIRST project to enable parallel RTHS at millisecond time scales; and the RT-Xen real-time virtualization research platform and the RTDS scheduler that is now part of the Xen open-source software distribution. Professor Gill has served as an Associate Editor for TCPS and Subject Area Editor for the Elsevier Journal of Systems Architecture. He has served in numerous other organizing and technical reviewing roles within the real-time systems research community, including: IEEE TCRTS Chair; IEEE TCRTS Vice-Chair; IEEE RTSS General Chair; ACM SIGBED Vice-Chair; IEEE RTSS Technical Program Committee Chair; IEEE TCRTS Treasurer and IEEE RTSS Finance Chair. 
\end{IEEEbiography}
\vspace{-15mm}

\begin{IEEEbiography}[{\includegraphics[width=1in,height=1.25in,clip,keepaspectratio]{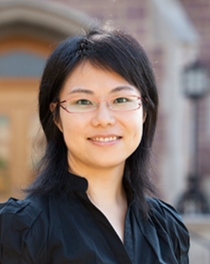}}]{Xuan Zhang}
is an Associate Professor in the Preston M. Green Department of Electrical and Systems Engineering at Washington University in St. Louis. She works across the fields of VLSI design, computer architecture, and cyber-physical systems and her research interests include hardware/software co-design for efficient machine learning and artificial intelligence, adaptive power and resource management for autonomous systems in analog/mixed-signal and physical domain.
Before joining Washington University, Dr. Zhang was a Postdoctoral Fellow in Computer Science at Harvard University. She received her BE degree in Electrical Engineering from Tsinghua University in China, and her MS and Ph.D. degrees in Electrical and Computer Engineering from Cornell University. Dr. Zhang is the recipient of NSF CAREER Award in 2020, AsianHOST Best Paper Award in 2020, DATE Best Paper Award in 2019, and ISLPED Design Contest Award in 2013, and her work has also been nominated for Best Paper Awards at ASP-DAC 2021, DATE 2019 and DAC 2017.
\end{IEEEbiography}

\end{document}